\documentclass[12pt]{iopart}

\usepackage{iopams}  
\usepackage{graphicx}
\begin{document}

\title[CCB Mass-Shell Model to Other GW Events]{Comparing a Compact-Binary Mass-Shell Model with Select Observed Gravitational Waves}

\author{Noah M. MacKay%\,\orcidlink{0000-0001-6625-2321}
\\ ORCID ID: 0000-0001-6625-2321}

\address{Institut für Physik und Astronomie, Universit\"at Potsdam,\\ Karl-Liebknecht-Str. 24/25, 14476 Potsdam, Germany}
\ead{noah.mackay@uni-potsdam.de}
\vspace{10pt}
\begin{indented}
\item[]\today
\end{indented}

\begin{abstract}
In a recent work, coalescing compact binaries (CCBs) were modeled as a rotating and contracting compact mass shell, providing an alternative effective representation to the state-of-the-art effective-one-body approach. Using a variational methodology, the Laplace-Beltrami formulation of the Ricci tensor was applied to a Kerr metric Ansatz, and the corresponding energy density $T_{00}$ of the CCB mass shell was obtained via the Einstein field equations. At the time of coalescence $t_C$, the resulting surface energy depends on the reduced mass $\mu$, the symmetric mass ratio $\alpha$, and the normalized orbital spin velocity of the CCB. In this work, we evaluate the radiated energy predicted by this variational approach for 45 select gravitational wave (GW) events from the O1--O4 runs, and compare these values with those inferred from observational catalogs, either directly or via the total-minus-remnant mass difference. For 38 of the 45 events analyzed, the predicted radiated energies agree with observationally inferred values, with 1:1 ratios spanning from $0.828$ to $0.997$ (mean $0.942$, median $0.955$). Three events exhibit ratios in the range $0.721\sim0.779$, one event yields a ratio of $0.466$, and for the remaining events the radiated energy is either unconstrained or inaccessible due to undocumented total-minus-remnant mass differences. These results indicate that the analytical approximation captures, for the most part, the leading-order energy scaling of compact binary mergers, while also suggesting clear avenues for further systematic improvement, including incorporating post-Newtonian corrections due to e.g. a non-zero eccentricity or combined tidal deformability.
\end{abstract}

\submitto{\CQG}
\maketitle

\section{Introduction} \label{intro}

Since their discovery on September 14, 2015, gravitational waves (GWs) have been routinely observed by the LIGO-Virgo-KAGRA (LVK) collaboration \cite{GWOSC, LIGOScientific:2018mvr, LIGOScientific:2021usb, KAGRA:2021vkt, LIGOScientific:2025slb}, now called the International Gravitational Wave Network (IGWN).  The dominant astrophysical sources of GWs have been well understood to be coalescing compact binaries (CCBs), whose leading-order radiation is quadrupolar in nature. By linearizing the Einstein field equations (EFEs) in a weak field limit and modeling the source as a compact binary with two point-mass constituents, one obtains the traceless-transverse (TT) gauged spatial waveforms, e.g. for simple circular orbits: 
\begin{equation}\label{gwsourced}
 h^\mathrm{TT}_{ij}=-\frac{4G\mu L^2\Omega^2}{D}\exp(i\,2\Omega t)\varepsilon^\mathrm{TT}_{ij},
\end{equation} 
where $D$ is the luminosity distance between the observer and the source, $\mu=m_1m_2/M$ is the reduced mass of the binary ($M=m_1+m_2$ is the total mass), $L$ is the binary separation, $\Omega$ is the orbital frequency,  $h_+\propto\cos(2\Omega t)$ and $h_\times\propto\sin(2\Omega t)$ via the $3\times3$ TT-gauged matrix $\varepsilon^\mathrm{TT}_{ij}$, and the strain amplitude is distinctly defined in source-dependent values. In this work $c=1$ is adopted. 

However, realistic CCBs complicate the simple waveforms to incorporate dynamic post-Newtonian (PN) \cite{Blanchet:2013haa} and perturbative post-Minkowskian (PM) \cite{Damour:2016gwp} corrections, and beyond the PN regime CCBs undergo the inspiral-merger-ringdown (IMR) process. In these latter phases, the waveform intensifies with a dynamic frequency and amplitude enhancement, until reaching the maximum peak at the coalescence time $t_C$. Past coalescence, ignoring tidal deformations, the waveform dampens exponentially towards a zero flat-line.

Analytical efforts to describe CCB dynamics include the effective one-body (EOB) framework \cite{Buonanno:1998gg, Buonanno:2005xu}, in which the CCB is mapped onto an effective one-body problem to employ the reduced mass $\mu$ moving in a deformed background. In this description, the effective reduced-mass particle inspirals from large separations towards the total mass innermost stable circular orbital (ISCO) radius, until it plunges into the total mass horizon with radius $r_S=2GM$. The largeness of the symmetric mass ratio $\alpha:=\mu/M\in(0,1/4)$ strongly influences the longevity of nearly-circular orbital paths and how soon they transition into plunge paths \cite{Buonanno:1998gg}. The energetics of CCBs in the EOB framework are encoded in the EOB Hamiltonian $H_\mathrm{EOB}$ \cite{Buonanno:1998gg, Damour:2009zoi, Damour:2012mv}, which reads as \cite{Buonanno:1998gg}:
\numparts
\begin{eqnarray}
&H_\mathrm{EOB}=M\sqrt{1+2\alpha\left(\frac{H_\mathrm{eff}^\alpha}{\mu}-1\right)},\\
&\mathrm{where}\quad H_\mathrm{eff}^\alpha=\mu\sqrt{A_\alpha(r)\left[1+\frac{\mathbf{p}^2}{\mu^2}+\left(\frac{1}{B_\alpha(r)}-1\right)\frac{p_r^2}{\mu^2} \right]}\\
&\mathrm{with}\quad ds^2_\mathrm{eff}=-A_\alpha(r)dt^2+B_\alpha(r)dr^2+r^2d\Omega^2.
\end{eqnarray}
\endnumparts
Although the EOB framework can be systematically derived from the EFEs and augmented via PN and PM information, practical applications rely on numerical evaluation and calibration against high-precision numerical-relativity waveforms. The effective phase-space variables $\mathbf{p}$ and $p_r$ respectively encode tangential and radial motion of the reduced mass, and the metric functions $A_\alpha(r)$ and $B_\alpha(r)$ incorporate higher-order conservative dynamics as well as resummation effects. Despite its remarkable accuracy and central role in modern GW analysis, evaluating EOB dynamics and associating waveform families remains computationally (and temporary) expensive. This motivates the exploration of complementary, analytically tractable methods to find key measurables -- more specifically the total radiated energy at coalescence -- however at the trade-off of being an approximation.

\begin{figure*}[h!]
\centering
	\includegraphics[width=\textwidth]{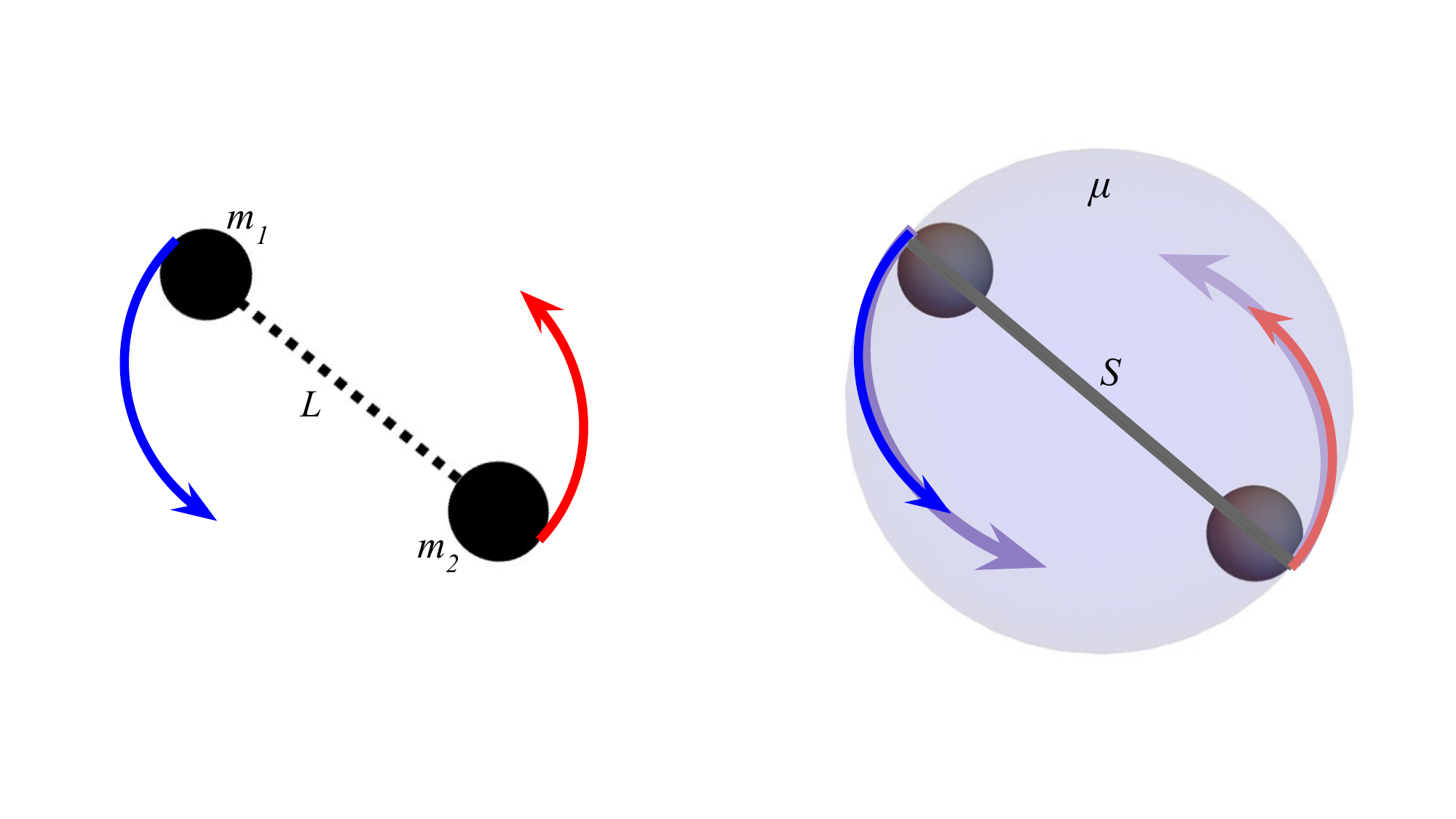}
\caption{\label{fig:mod} A visual comparison between an ideal two-body CCB (left; two orbiting masses $m_1$ and $m_2$ separated by length $L$) and the mass-shell representation (right; the shell of distributed mass $\mu$ and diameter $S$ via Eq. (\ref{rads}) is displayed with respect to the CCB seen on the left). Unlike the typical ``marble in the funnel" picture of EOB, the mass-shell representation on the right assumes the orbital frequency of the CCB as its angular (axial) frequency -- thus having an angular momentum $J=I\Omega$ --, and $S(t)$ shrinks with $L(t)$ until reaching their final states at time $t_C$.}
\end{figure*}

One such alternative is a hollow mass-shell model for CCBs, introduced in Ref. \cite{MacKay:2024qxj}. Rather than explicitly treating the dynamics of a reduced mass orbiting in an effective potential (i.e., a ``marble in a funnel" picture), the binary masses $m_1$ and $m_2$ are represented as an effective, reduced mass distribution confined to a hollow, spherical mass shell with a decreasing radius and an increasing angular frequency during its evolution. Coalescence ends in this picture at the ``innermost" shell radius of $\rho=2GM$, consistent with the EOB framework. A visual aid is given in Figure \ref{fig:mod}, in Section \ref{model}, where the mass-shell representation is compared with an ideal two-body CCB.

Essentially, this mass-shell representation is applicable for all stages in the inspiral-merger phases. As discussed in Ref. \cite{MacKay:2024qxj}, the effective and purely phenomenological model of the CCB as an evolving mass-shell allows one to map the compact binary's morphology from early inspiral to near-merger (i.e., from a ``wide" and slow-rotating mass-shell to a ``compact" and fast-rotating mass-shell). From this morphology in the mass-shell, one could extract e.g. low-frequency waveform profiles and the GW energy radiated during inspiral, naming only a couple projects proposed by the analytical initative of the Einstein Telescope (ET) collaboration \cite{ET:2025xjr}. The ringdown phase is excluded, as it describes the remnant object with final-state parameters (e.g. mass $M_\mathrm{f}$ and spin $\chi_\mathrm{f}$) that is best represented conventionally. However, more informative -- provided the frequency sensitivity of current-generation laser interferometers across the IGWN -- and relevant stages of inspiral-merger to apply the mass-shell representation rest in the late-inspiral / pre-merger stages, where the mass-shell encompasses (in a sense) the two objects, with two distinct horizons, as an effective ``outer horizon". 

One can view this mass-shell model as a reinterpretation of EOB, which enables us to effectively assume CCBs as a singular compact object. Therewith, we can make claims and assumptions consistent with previous EOB and CCB approaches, most importantly $D\gg L$ for very distant sources. It is worth repeating that this mass-shell representation, aside from the ``marble in the funnel" picture of conventional EOB, is meant to be phenomenological -- a useful tool to apply e.g. variational methods on the EFEs as a means to approximate the coalescence energy of radiated GWs \cite{MacKay:2024qxj}. In addition, one can follow Newton's two postulates for a hollow mass shell: respectively, the gravitational field outside the shell is equivalent to that of a point mass, and the gravitational field in the interior is nullified. This treats the generated GWs and their respective waveform as radiation emitted along the shell surface, somewhat analogous to Hawking radiation. We review the mass-shell model for comprehension in Section \ref{model}.

For the mass-shell model to determine the energy eigenvalue for GW coalescence, a methodology different from the conventional hierarchy of solving the EFEs\footnote{Normally, we start with an arbitrary metric and obtaining the Einstein tensor components to set equal to well-defined energy-momentum tensor components.} is needed.  For a rotating hollow shell, constructing a custom metric from the ground up and carrying out the full procedure would be a cumbersome task -- especially given that the objective is to uniquely define the energy density $T_{00}$ of our hollow mass shell, which gives the energy eigenvalue $E=T_{00}V$ when evaluated at the shell surface. A more effective alternative is to employ a method analogous to the quantum-mechanical variational method: we begin with an Ansatz metric $g_{\mu\nu}$ that is already a known solution to the EFEs, and then apply differential operations inherent in $G_{\mu\nu}$ in order to extract its effective components, thereby obtaining effective expressions for the corresponding $T_{\mu\nu}$ components. Since the mass-shell model qualitatively mimics a Kerr-like object, it adopts the Kerr metric Ansatz, and through the Laplace-Beltrami formulation of the Ricci tensor, one obtains (under simple assumptions) the coalescence energy eigenvalue:
\begin{equation}\label{esimp}
E(t_C)\simeq\frac{\pi}{6}\alpha\mu \left(1-5\beta_C^2\right),
\end{equation}
where $\beta_C=v_C/c$ in SI units, briefly recovering $c$, is the speed ratio depending on the normalized rotation speed of the CCB at coalescence, with $v_C$ to take the expected form $v_C=GMf_\mathrm{GW,peak}$ (that is, the peak orbital frequency multiplied by the total mass horizon radius). Its derivation, together with a detailed reintroduction of the Laplace-Beltrami formalism, is provided in Section \ref{shelen}.

Two representative examples: GW150914 \cite{LIGOScientific:2016aoc} and GW170817 \cite{LIGOScientific:2017vwq}; were selected mainly for their distinct CCB types -- respectively binary black hole (BBH) and binary neutron star (BNS) -- to test the approximation's invariance across source types by computing the expected energy and comparing it with cataloged energy values. Using central parameter values (low-spin prior for GW170817) reported in the respective GWTC catalogs \cite{LIGOScientific:2018mvr, LIGOScientific:2021usb}, Eq. (\ref{esimp}) yields anticipated energies of $2.075M_\odot$ and $0.0885 M_\odot$, again with $c=1$. These are to be compared with the corresponding cataloged values $E_\mathrm{rad}=3.1^{+0.4}_{-0.4}M_\odot $ and $E_\mathrm{rad}\geq 0.04 M_\odot $ \cite{LIGOScientific:2018mvr}. 

The scaling discrepancy between the model approximation and the cataloged values originates from the simplifying assumptions made in deriving Eq. (\ref{esimp}). As demonstrated in Ref. \cite{MacKay:2024qxj}, a more comprehensive treatment using the Kretschmann scalar as a curvature surrogate, $R_\mathrm{eff} = -\sqrt{K}/6$, yields significantly improved quantitative agreement with cataloged merger energies -- for example, $E(t_C)\simeq 3.26M_\odot $ and $E(t_C)\simeq 0.14M_\odot $ for GW150914 and GW170817, respectively. In the present work, we revisit the derivation of Eq. (\ref{esimp}), but now employ the Kretschmann-scalar surrogate and explicitly discuss its physical significance. We extend the comparison between the model predictions and observations across, in total, 45 of the 368 cataloged GW events as of 30 November 2025; these include our previous representative examples. The results are presented in Section \ref{sign}, followed by a discussion in Section \ref{disc} and concluding statements in Section \ref{concl}.

\section{Reviewing the CCB Mass-Shell Model} \label{model}

For both stable classical binaries and CCBs, the reduced mass $\mu$ simplifies two-body dynamics into a singular effective system. For CCBs specifically, the binary may be assumed to behave as a singular object when viewed from far away, i.e., at a very large luminosity distance $D$. However, rather than the reduced mass behaving like a marble spiraling down a funnel, it is instead distributed over a rotating and contracting mass shell. While undergoing coalescence, the rate of change in the separation $-\dot{L}$ into a final separation $L(t_C)$ relates to a mass shell's contracting  diameter with the rate of change $-\dot{S}$ and a final diameter measure $S(t_C)$. The overdot resembles a time differential: $^\bullet\equiv d/dt$. When integrated over the timelapse $t'\in[t,\,t_C]$, where $t$ is dynamic and $t_C$ is fixed, we yield:
\begin{equation}\label{seps}
-S(t_C)+S(t)=-L(t_C)+L(t).
\end{equation}
In Eq. (\ref{seps}), the CCB separation and shell diameter at the coalescence time $t_C$ are fixed values. Adopting EOB convention by imposing coalescence ends when the masses make contact: $L(t_C)=r_1+r_2$, under the shell diameter that identifies the total mass horizon diameter: $S(t_C)=4GM$, we define:
\begin{equation}\label{rads}
S(t)=L(t)-r_1-r_2+4GM.
\end{equation}
In this model, the EOB picture is reinterpreted such that the system is viewed instead as a shrinking mass shell with the constant measure $\mu$ and contracting radius $\rho(t)=S(t)/2$, until reaching the ``innermost'' shell that is the total mass horizon. 

\subsection{Mass-Shell Waveforms}

Even in the CCB mass-shell model, the respective TT-gauged spatial waveform of this mass shell model is conventionally quadrupolar:
\begin{equation}\label{wave}
h^\mathrm{TT}_{ij}=\frac{2G}{D}\ddot{Q}_{ij}^\mathrm{TT},
\end{equation}
where $Q_{ij}$ is the quadrupole moment tensor. For a CCB treated as an effective compact object rotating at the dynamic orbital velocity $\Omega=\Omega(t)$,  the quadrupole moment is proportional to the point mass moment of inertia $I=\mu \rho^2$, e.g.:
\numparts
\begin{eqnarray}\label{qij}
& Q_{ij}=\frac{1}{4}\mu S^2\left(c_ic_j-\frac{1}{3}\delta_{ij}\right),\\
&\mathrm{with}\quad\vec{c}=(\cos(\Omega t),\,\sin(\Omega t),0),\label{vecs}
\end{eqnarray}
\endnumparts
if we assume (nearly-)circular orbits. One introduces eccentric/elliptical orbits at this stage of the derivation, as one would set e.g. $\vec{c}=2(r_+\cos(\Omega t)+u,\,r_-\sin(\Omega t)+w,0)/S$ to encode the semi-major and semi-minor radii and the center displacement $(u,w)$, or alternatively by allowing a PN-like expansion in $\Omega$ to include an off-circular correction. Nonetheless, via Eq. (\ref{wave}) and using Eqs. (\ref{qij}) and (\ref{vecs}), we yield the mass-shell source, (nearly-)circular GW profile readily in the TT-gauge, also considering time variation in the CCB separation $S=S(t)$ and angular velocity $\Omega$
\numparts
 \begin{eqnarray}\label{hplus}
&h_{+}^{\mathrm{TT}}=-\frac{G\mu S^2}{D}\Big(\sin(2\Omega t)\left(\frac{2\dot{S}}{S}\left(\Omega +t\dot{\Omega}\right)+\dot{\Omega}+\frac{t}{2}\ddot{\Omega}\right)\\\nonumber
&\quad\quad\quad\quad\quad\quad\quad -\cos(2\Omega t)\left(\frac{\dot{S}^2}{2S^2}+\frac{\ddot{S}}{2S}-\left(\Omega+t\dot{\Omega}\right)^2\right)+\frac{\dot{S}^2}{2S^2}+\frac{\ddot{S}}{2S}\Big),\\ \label{hcross}
&h_{\times}^\mathrm{TT}=-\frac{G\mu S^2}{D}\Big(\sin(2\Omega t)\left(\frac{\dot{S}^2}{2S^2}+\frac{\ddot{S}}{2S}-\left(\Omega+t\dot{\Omega}\right)^2\right)\\\nonumber
&\quad\quad\quad\quad\quad\quad\quad +\cos(2\Omega t)\left(\frac{2\dot{S}}{S}\left(\Omega +t\dot{\Omega}\right)+\dot{\Omega} +\frac{t}{2}\ddot{\Omega}\right)\Big).
\end{eqnarray}
\endnumparts
If we assume, for early inspiral stages, large seperations such that $S\approx L$ and perfectly circular orbits, where all rates of change are zero, we recover the wave profiles nearly identical to Eq. (\ref{gwsourced}), however scaled by $1/4$ due to our choice of quadrupole geometry. To restore the factor of $4$, one can instead adopt the standard two-point-mass quadrupole moment, which bypasses the moment of inertia in favor of $m_1r_1^2+m_2r_2^2=\mu L^2$. In this case, if one chooses not to approximate $S\approx L$ and keep $S$ in the waveforms, one would then rewrite $L$ by rearranging Eq. (\ref{rads}).

 However, unlike the simple circular wave model in which both separation and angular frequency are fixed, these quantities are dynamic across coalescence while conserving angular momentum $J=I\Omega$. Imposing the conservation law $\dot{J}=0$ for the mass-shell geometry yields differential relations linking the rates of change in $S$ and $\Omega$, and leads to waveform expressions written entirely in terms of time derivatives of e.g. $\Omega$, c.f. Ref. \cite{MacKay:2024qxj}:
\numparts
 \begin{eqnarray}\label{hp2}
&h_{+}^{\mathrm{TT}}=-\frac{G\mu S^2}{D}\Big(\sin(2\Omega t)\left(-\frac{\dot{\Omega}}{\Omega}\left(\Omega +t\dot{\Omega}\right)+\dot{\Omega}+\frac{t}{2}\ddot{\Omega}\right)\\\nonumber
&\quad\quad\quad\quad\quad\quad\quad -\cos(2\Omega t)\left(-\frac{\ddot{\Omega}}{4\Omega}+\frac{\dot{\Omega}^2}{4\Omega^2}-\left(\Omega+t\dot{\Omega}\right)^2\right)-\frac{\ddot{\Omega}}{4\Omega}+\frac{\dot{\Omega}^2}{4\Omega^2}\Big),\\\label{hc2}
&h_{\times}^\mathrm{TT}=-\frac{G\mu S^2}{D}\Big(\sin(2\Omega t)\left(-\frac{\ddot{\Omega}}{4\Omega}+\frac{\dot{\Omega}^2}{4\Omega^2}-\left(\Omega+t\dot{\Omega}\right)^2\right)\\\nonumber
&\quad\quad\quad\quad\quad\quad\quad +\cos(2\Omega t)\left(-\frac{\dot{\Omega}}{\Omega}\left(\Omega +t\dot{\Omega}\right)+\dot{\Omega} +\frac{t}{2}\ddot{\Omega}\right)\Big).
\end{eqnarray}
\endnumparts
These waveform expressions, provided late insprial stages during times $t<t_C$ and the near-merger stage at time $t\sim t_C$, are more relevant to utilize as useful phenomenological tools to describe GWs from the mass-shell representation. More explicit expressions rest in one's choice in e.g. defining an Ansatz for $\Omega$ and its derivatives systematically.

Given that the orbital expressions are directly coupled to the wave profiles, and thus govern the evolution of the waveform envelope during coalescence, one must also consider any post-merger dynamics to complete the waveform. For times $t>t_C$, the GW signal transitions into its ringdown phase, which may be modeled using quasi-normal modes and piecewise stiching, e.g. with a Gaussian-like damping inspired by stellar collapse and post-merger relaxation \cite{Buonanno:2005xu, Gundlach:1993tp}. In the general sense, taking Eqs. (\ref{hp2}) and (\ref{hc2}) as the inspiral-merger basis, the complete IMR waveform may be expressed as a linear expansion in sine and cosine profiles, each scaled by a time-dependent, polarization-specific envelope function $\mathfrak{E}(t)$ encapsulating the dynamics:
\begin{equation} \label{pieces}
h_{+/\times}^\mathrm{TT}(t)=-\frac{G\mu S^2}{D}\left[\cos(2\Omega t) \mathfrak{E}_{+/\times}^\mathrm{cos}(t)+\sin(2\Omega t)\mathfrak{E}_{+/\times}^\mathrm{sin}(t)\right].
\end{equation}
E.g., for the simple wave approximation whereby $\dot{\Omega},~\ddot{\Omega}\rightarrow0$, $\mathfrak{E}(t)$ reduces uniquely into a constant (either $\Omega^2$ or 0, depending on polarization) for all $t$, and the waveform is scaled only by its amplitude, similar to Eq. (\ref{gwsourced}). In contrast, the fully dynamic IMR amplitude is encoded in a characteristic piecewise structure of $\mathfrak{E}(t)$; for times $t\leq t_C$, this is  given by the complete expressions in Eqs. (\ref{hp2}) and (\ref{hc2}). For times $t>t_C$, this transitions into a Gaussian-like damping form, motivated by Ref. \cite{Gundlach:1993tp}, resulting to the asymptotic level-out to zero. For BNSs in particular, the envelope function $\mathfrak{E}(t)$ acquires additional tidal contributions entering through the PN corrections to $\Omega$ up to merger, followed by non-trivial post-merger scaling prior to a possible BH collapse. 

\subsection{Mass-Shell Surface Energy}\label{shelen}

While it is conventional to extract the energy density (and energy flux density) of GWs via the invariant combination $h_+^2+h_\times^2$, using either Eqs. (\ref{hp2}) and (\ref{hc2}) together or Eq. (\ref{pieces}) generally, a more heuristic approach was employed in Ref. \cite{MacKay:2024qxj}, wherein the relevant quantity was obtained through a variational treatment of the EFEs. The EFEs describe gravity as the dynamic response of spacetime geometry to matter-energy sources:
\begin{equation}\label{efes}
G_{\mu\nu}:=R_{\mu\nu}-\frac{1}{2}Rg_{\mu\nu}=8\pi G\,T_{\mu\nu}.
\end{equation}
The Einstein tensor $G_{\mu\nu}$ on the left-hand side encodes the geometric responses caused by the source $T_{\mu\nu}$ via the Ricci $R_{\mu\nu}$ and metric $g_{\mu\nu}$ tensors and the Ricci scalar $R=g^{\mu\nu}R_{\mu\nu}$. The additive cosmological contribution $\Lambda g_{\mu\nu}$ is neglected here. The energy-momentum tensor $T_{\mu\nu}$ on the right-hand side defines our source; notable examples include vacuum $T_{\mu\nu}=0$ and a perfect fluid $T_{\mu\nu}=(\epsilon+p)u_\mu u_\nu+pg_{\mu\nu}$. For GW sources, which are inherently quadrupolar, the effective source term $T_{\mu\nu}$ is treated as non-vanishing, independent of whether the CCB is a BBH, BNS, or BH-NS system. Under linearization about a flat background and assuming a non-trivial source, this leads to the wave equation given in Eq. (\ref{wave}). However, as the central objective of the effective CCB mass-shell picture is the surface energy eigenvalue rather than the spacetime perturbations themselves, we need to consider a different methodology.

On one hand, known solutions to the EFEs, e.g. Schwarzschild \cite{Schwarzschild:1916uq, Schwarzschild:1916ae}, Tolman-Oppenheimer-Volkoff \cite{Tolman:1939jz, Oppenheimer:1939ne}, and Kerr \cite{Kerr:1963ud, Boyer:1967, Chandrasekhar:1985kt}, were obtained by first specifiying a well-defined energy-momentum source, either vacuum \cite{Schwarzschild:1916uq, Schwarzschild:1916ae, Kerr:1963ud, Boyer:1967} or a perfect fluid \cite{Tolman:1939jz, Oppenheimer:1939ne}. On the other hand, the present objective is to determine the effective energy density $T_{00}$ for a CCB modeled as a rotating and contracting mass shell. To remain consistent with the interpretation of CCBs as energetic sources of gravitational radiation, this quantity cannot vanish. Consequently, rather than beginning with an explicit form of $T_{\mu\nu}$, we instead begin with a well-defined geometric Ansatz for the spacetime configuration. Although this inversion of the usual procedure may appear counterintuitive, it becomes tractable under a variational framework.

The variational method (also called the approximational method) is commonly used in perturbative quantum mechanics to estimate the energy eigenvalue of an otherwise intractable Hamiltonian. This is achievable by first assuming an Ansatz wavefunction, then evaluating the expectation value of the Hamiltonian to obtain an energy functional, and lastly minimizing the functional to obtain the anticipated ground state energy. Here, we follow a similar logic, as initially done in Ref. \cite{MacKay:2024qxj}: we choose a well-defined metric Ansatz that best describes the system -- one that is already a known solution to the EFEs --, on which we apply differential operators inherent in the Christoffel symbols $\Gamma^\alpha_{\mu\nu}$ and the Ricci tensor, and define $G_{\mu\nu}\propto T_{\mu\nu}$ component-wise, focusing exclusively on $G_{00}\propto T_{00}$. 

Schematically, the Ricci tensor has the structure $R\sim\partial\Gamma+\Gamma\Gamma$ (suppressing the indices), while the Christoffel symbols satisfy $\Gamma\sim g^{-1}\partial g$. Therefore, we claim that the Ricci tensor can be effectively defined as a Laplace-like operator acting on the metric tensor: $R\sim \nabla(g^{-1}\partial g)\sim\nabla^2g$, where $\nabla\sim(\partial+\Gamma)$ is the covariant derivative (see footnote\footnote{One might be wary of this definition, given metric compatibility $\nabla_\mu g^{\mu\nu}=0$. However, it is shown that the covariant Laplacian: the Laplace-Beltrami operator, is coordinate-dependent and provides non-trivial results when acted on specific metric tensor components.}). This interpretation is consistent with Lemma 3.32 from Chow and Knopf \cite{Chow:2004}, which states in leading order:
\begin{equation}\label{ricci}
R_{\mu\nu}\simeq-\frac{1}{2}\Delta^\mathrm{LB} g_{\mu\nu}\,\left(+~\mathrm{lower~order~terms}\right).
\end{equation}
The operator $\Delta^\mathrm{LB}$ is the Laplace-Beltrami operator, which for a well-defined metric with a non-zero determinant may be written in the Christoffel symbol-free form as:
\begin{equation}\label{lb}
\Delta^\mathrm{LB}=\frac{1}{\sqrt{-g}}\,\partial_\alpha\left(\sqrt{-g}g^{\alpha\beta}\partial_\beta \right),
\end{equation}
where $\sqrt{-g}=\sqrt{-\mathrm{det}(g_{\mu\nu})}$ and $g^{\mu\nu}$ is the inverse metric. From Eq. (\ref{efes}), it follows that $G_{00}$ depends on $R_{00}$, $R$, and $g_{00}$, and through Eq. (\ref{ricci}) $R_{00}\propto g_{00}$. In this Laplace-Beltrami formulation, the effective energy density is defined approximately however essentially through $g_{00}$, dropping the lower order terms in Eq. (\ref{ricci}):
\begin{equation}\label{endens}
T_{00}\approx -\frac{1}{8\pi G}\left(\frac{1}{2}\Delta^\mathrm{LB} +\frac{1}{2}R \right)g_{00}.
\end{equation}
One can see that, as $g_{00}$ is negative under the $(-,+,+,+)$ metric signature, this approximation yields a positive calculation. 

Both $\Delta^\mathrm{LB}$ in Eq. (\ref{lb}) and the Ricci scalar $R$ depend on one's choice of Ansatz metric. For a CCB mass shell observed from a luminosity distance $D\gg S$, we can assume the system to resemble a spinning, compact object. Therefore, our choice of Ansatz metric is the Kerr metric \cite{Kerr:1963ud, Boyer:1967, Chandrasekhar:1985kt} in Boyer-Lindquist coordinates, written here for a given spinning mass of measure $m$:
\begin{eqnarray}\label{kermet}
ds^2=&-\left(1-\frac{2Gm r}{\Sigma} \right)dt^2+\frac{\Sigma}{\Delta}dr^2+\Sigma d\theta^2\\\nonumber
&+\left(r^2+a^2+\frac{2Gm ra^2}{\Sigma}\sin^2\theta \right)\sin^2\theta d\varphi^2-\frac{4Gm ra\sin^2\theta}{\Sigma} d\varphi dt,
\end{eqnarray}
where $a=J/m$ is a length-scale spin parameter, and
\numparts
\begin{eqnarray}
&\Sigma=r^2+a^2\cos^2\theta,\\
&\Delta=r^2-2Gm r+a^2.
\end{eqnarray}
\endnumparts
To apply the Kerr metric to the CCB mass-shell model, we simply substitute $m\rightarrow\mu$ for the mass measure. The metric determinant under the Kerr metric is given as $\mathrm{det}(g_{\mu\nu})=-\Sigma^2\sin^2\theta$, and since the metric elements depend only on the coordinates $r$ and $\theta$, the resulting Laplace-Beltrami operator is compact:
\begin{equation}\label{lbkerr}
\Delta^\mathrm{LB}_\mathrm{Kerr}=\frac{1}{\Sigma \sin\theta}\left[ \sin\theta \partial_r\left(\Delta  \partial_r\right)+\partial_\theta\left(\sin\theta \partial_\theta\right) \right].
\end{equation}

The Ricci scalar of the Kerr metric, however, is $R=0$. While the Kerr metric is conventionally a vacuum solution of the EFEs, for which $R_{\mu\nu}=0$, the coordinate-dependent Laplace-Beltrami formulation of the Ricci tensor, i.e. via Eqs. (\ref{ricci}) and (\ref{lbkerr}), enables one to obtain effective, non-zero expressions of the curvature components for any well-defined metric Ansatz via that metric's respective components. This includes the Kerr metric, as demonstrated in Ref. \cite{MacKay:2024qxj}, again as long as we consider the distinctive and individual metric components. Generally, if we do not distinctly define the metric Ansatz and the choice of coordinates, $\Delta^\mathrm{LB}=\nabla_\alpha \nabla^\alpha$ canonically, which renders $R_{\mu\nu}\propto \nabla_\alpha \nabla^\alpha g_{\mu\nu}=0$ by default of metric compatibility. One may claim, in this sense, that the vacuum nature of e.g. Kerr spacetime is recovered, albeit this holds for any metric. A detailed consistency check of satisfying the Bianchi identities in the Laplace-Beltrami formalism is offered in Appendix A in Ref. \cite{MacKay:2024qxj}, whereby the lower-order terms in Eq. (\ref{ricci}) are explicitly defined and proved to be essential.

In this coordinate-dependent Laplace-Beltrami formalism, retrieving e.g. the energy density $T_{00}$ is made convenient. One first derives $R_{00}\propto\Delta ^\mathrm{LB}g_{00}$ given Eqs. (\ref{kermet}) and (\ref{lbkerr}), yielding a contribution to a non-zero, effective evaluation of $G_{00}$. Therefore, $T_{00}$ is an effective, phenomenological expression intended to be a useful tool in our variational approach. Via Eq. (\ref{ricci}), one can straightforwardly compute $R_{00}$ to be a non-zero contribution:
\begin{eqnarray}\label{r00}
R_{00}=\frac{G^2\mu^2}{2\Sigma^4}\Big(&a^4\left(\frac{3}{2}+\frac{1}{2}\cos(4\theta)\right)\\\nonumber
&+4a^2\left(\cos(2\theta)\left(\frac{a^2}{2}-3r^2\right)-3r^2\right)+4r^4 \Big).
\end{eqnarray}
The physically relevant range in $r$, while normally covering the full radial range $\in[0,\infty)$, here extends across $\in[\rho,\infty)$, cutting off at the shell radius $\rho$. This is analogous to Birkhoff's theorem regarding external vacuum metrics for spherically-symmetric bodies.
  
In approaching the Ricci scalar $R$, we can either: (i) set $R=0$ in accordance with the Kerr metric Ansatz, (ii) heuristically construct $R$ using a Laplace-Beltrami trace over the metric: $R\propto g^{\mu\nu}\Delta^\mathrm{LB} g_{\mu\nu}$, or (iii) define an effective Ricci scalar by leveraging the curvature content of the non-vanishing Kretschmann scalar: $K=R_{\alpha\beta\mu\nu}R^{\alpha\beta\mu\nu}$ \cite{dInverno:1992gxs}. Choosing, e.g., option (i) implies that Eq. (\ref{endens}) only depends on a reduced Einstein tensor $G_{00}\approx R_{00}$ via Eq. (\ref{r00}). If one makes this choice due to its straight-forwardness, we Taylor expand Eq. (\ref{r00}) under a small spin-parameter $a$ (recalling that the final spin parameter of CCBs is typically $a<1$ \cite{LIGOScientific:2018mvr, LIGOScientific:2021usb, KAGRA:2021vkt, LIGOScientific:2025slb}, see footnote\footnote{In the GWTC catalogs, the dimensionless spin parameter is defined as $\chi\equiv a/(GM)$, and it is the conventionally reported quantity for final-spin measurements. Since $a=\rho\beta$ on the mass-shell surface, where $\beta$ is the rotational velocity ratio across coalescence, we obtain $a_C=\rho_C\beta_C\lesssim4GM/5$ at $t=t_C$. Therefore, at merger, $\chi_C=a_C/(GM)\lesssim4/5$, which is consistent with the observed clustering of final dimensionless spins in the range $\chi\sim0.6-0.8$ across GWTC events. Also, this indeed satisfies the Kerr bound $\chi<1$.}) and integrate over the polar angles $\theta\in[0,\pi]$ (an analytical consequence of the model's shell geometry, whereby the inclination angle $\iota\in[0,\pi]$ serves as the effective polar angle). This yields, via Eq. (\ref{endens}):
\begin{equation} \label{endens1}
T_{00}\simeq \frac{G\mu^2}{4r^4}\left(1-5\frac{a^2}{r^2} \right),
\end{equation}
and consequently the surface energy $E=T_{00}V$, evaluated at $r=\rho=S/2$. This further simplifies the spin term such that $a^2/\rho^2=\beta^2$, where $\beta$ encodes the normalized rotational velocity of the CCB. We can define this surface energy both generally and at the time $t=t_C$:
\begin{equation} 
E\simeq \frac{\pi}{3}\frac{G\mu^2}{\rho}\left(1-5\beta^2 \right)\quad\Rightarrow\quad E(t_C) \simeq \frac{\pi}{6}\frac{\mu^2}{M}\left(1-5\beta_C^2\right).
\end{equation}
This recovers Eq. (\ref{esimp}), which we know underestimates the GWTC energy values of our representative cases. If we assume the representative cases to be generalizations for all possible cases, Eq. (\ref{esimp}) would underestimate all possible examples of cataloged GW events. This implies that $G_{00}$ must be systematically enhanced beyond the $R_{00}$ contribution alone, i.e., $-R$ must be non-zero and contribute additively in Eq. (\ref{endens}).

\subsection{The Surrogate for the Ricci Scalar}

The discrepancy between Eq. (\ref{esimp}) and the cataloged energy values for e.g. GW150914 and GW170817 (see Section \ref{intro}) originates from modeling assumptions. Assuming that both the variational methodology and the Kerr-like metric Ansatz remain valid, the underestimated calculations suggest the need for an enhancement via a non-zero effective Ricci scalar contribution (see Eq. \ref{endens}). In Ref. \cite{MacKay:2024qxj}, the Ricci scalar was initially computed via the Laplace-Beltrami contraction, $R\propto g^{\mu\nu}\Delta^\mathrm{LB} g_{\mu\nu}$, but the resulting expression diverged upon integration over the polar angle $\theta\in[0,\pi]$. A heuristic, yet geometrically motivated, alternative is to leverage the non-zero Kerr-metric Kretschmann scalar \cite{Henry:1999rm, Visser:2007fj} (written here with the reduced mass as the mass measure):
\begin{equation}\label{kretsch}
K_\mathrm{Kerr}=\frac{48G^2\mu^2}{\Sigma^6}\left(15\left(a^4r^2\cos^4\theta-a^2r^4\cos^2\theta\right) +r^6-a^6\cos^6\theta \right)
\end{equation}
as a surrogate for the Ricci scalar. The physically relevant range in $r$ is also $\in[\rho,\infty)$, as it is for Eq. (\ref{r00}). However, constructing an effective Ricci scalar directly from the Kretschmann scalar is not immediately intuitive. First and foremost, we restore dimensionality by asserting $R_\mathrm{eff}\propto\sqrt{K}$, as both carry curvature dimensions of $[\mathrm{length}]^{-2}$. 

Suppose we define $R_\mathrm{eff}=-\sqrt{K}$, purely as a heuristics exercise, for the Ricci scalar in Eq. (\ref{endens}). This is based on the earlier statement that $-R$ must be non-zero and contribute additively to the Ricci component minimum. We straight-forwardly define a systematically-enhanced, effective Einstein tensor component $G_{00}$ by implementing $K_\mathrm{Kerr}$ via Eq. (\ref{kretsch}) and $g_{00}$ in Eq. (\ref{kermet}). Doing so provides a polynomial in powers of $G$, i.e., a PM expansion \cite{Damour:2016gwp}:
\begin{eqnarray}
G_{00}\simeq&\frac{G^2\mu^2}{2\Sigma^4}\left(4r^4+4a^2(\dots)+a^4(\dots) \right)\\\nonumber
&-\sqrt{3(r^6+(\dots)-a^6\cos^6\theta)}\left(\frac{2G\mu}{\Sigma^3}-\frac{4G^2\mu^2r}{\Sigma^4} \right),
\end{eqnarray}
where higher order angular terms in $a^2$ and $a^4$, and the square root terms, have been abbreviated for brevity, c.f. Ref. \cite{MacKay:2024qxj}. As like before with $R_{00}$ in the simple case scenario, $G_{00}$ must be integrated over the polar angle $\theta\in[0,\pi]$ in order to extract a full-sphere value. However, angular integration over $\sqrt{K}$ via Eq. (\ref{kretsch}) proves to be a challenge, given angular terms inside the square root. While substitution techniques such as a Weierstraß transformation could be useful to tame the integral, we instead recall that the final spin parameter of CCBs is typically $a<1$ \cite{LIGOScientific:2018mvr, LIGOScientific:2021usb, KAGRA:2021vkt, LIGOScientific:2025slb}. That is, across all stages of coalescence, the spin parameter remains well below unity. This physical constraint motivates a Taylor expansion of $G_{00}$ in small $a$, truncated at quadratic order to capture a leading spin correction. Afterwards, then we integrate over $\theta$. Carrying out this procedure yields
\begin{eqnarray}\label{goo}
\Rightarrow G_{00}=&\frac{2\pi G^2\mu^2}{r^4}-\frac{10\pi G^2\mu^2}{r^4}\frac{a^2}{r^2}-\frac{2G\mu\sqrt{3}}{r^3}+\frac{21\pi G\mu\sqrt{3}}{r^3}\frac{a^2}{r^2}\\\nonumber
&+\frac{4\pi G^2\mu^2\sqrt{3}}{r^4}-\frac{23\pi G^2\mu^2\sqrt{3}}{r^4}\frac{a^2}{r^2}.
\end{eqnarray}
The first two terms are the ``small $a$'' polar-integrated $R_{00}$ contributions from Eq. (\ref{r00}), which enter at 2PM order. Discarding the 1PM terms by choice, Eq. (\ref{goo}), and with it $T_{00}$, simplifies to a modified zeroth-order Newtonian term and second-order spin correction compared to Eq. (\ref{endens1}):
\begin{equation}\label{too}
\Rightarrow\quad T_{00}\simeq 1.116\frac{G\mu^2}{r^4}\left(1-5.582\frac{a^2}{r^2} \right).
\end{equation}

Through the respective mass-shell surface energy:
\numparts
\begin{eqnarray} 
&E\simeq 4.675\frac{G\mu^2}{\rho}\left(1-5.582\beta^2 \right)\\
& \Rightarrow\quad E(t_C)=2.337\frac{\mu^2}{M}\left(1-5.582\beta_C^2 \right),
\end{eqnarray}
\endnumparts
it is evident that the anticipated values are systematically overestimated compared to Eq. (\ref{esimp}). This same conclusion of an overestimated approximation was made previously in Ref. \cite{MacKay:2024qxj}; the approximated energy for e.g. GW150914, using center parameter values, was found to be $E(t_C)\simeq 9.227 M_\odot $, well above the cataloged value. To provide physical reason behind this geometric overshoot, we recall the identity for the Kretschmann scalar:
\begin{equation}\label{kid}
K=C_{\alpha\beta\mu\nu}C^{\alpha\beta\mu\nu}+2R_{\alpha\beta}R^{\alpha\beta}-\frac{1}{3}R^2,
\end{equation}
where the first two terms are, respectively, the Weyl and Ricci curvature contributions. In vacuum solutions, the Kretschmann scalar is entirely sourced by the Weyl tensor, given the vanishing Ricci contributions. Without moderation, substituting $\sqrt{K}$ directly as an effective scalar injects a curvature scale that exceeds the quadrupolar radiation regime, leading to an overestimation of the coalescence energy. To solve this problem, we must manually insert a correcting calibrator to regulate this geometric overshoot, which we recall from Ref. \cite{MacKay:2024qxj} and from Section \ref{intro} that it is $1/6$. Some might accuse this as an ad hoc guess, but there is geometric significance. We first consider the $1/2$ factor inherent in Eq. (\ref{ricci}), which would have persisted in full variational calculations:
\begin{equation}
R\simeq-\frac{1}{2}\left(g^{\mu\nu}\Delta^\mathrm{LB} g_{\mu\nu}\right) \quad\Rightarrow\quad R_\mathrm{eff}\simeq-\frac{1}{2}\left(\tilde{\lambda}\sqrt{K} \right),
\end{equation}
where $\tilde{\lambda}$ is a proportionality coefficient. Next, we set the condition that $\tilde{\lambda}\in\mathbb{N}$, so that the quantity remains geometrically significant. However, we constrain that $\tilde{\lambda}<1$; the reason being that the initial, overshot case was the effective case whereby $\tilde{\lambda}=2$. As a result, we introduce a reciprocal proportionality constant $\tilde{\lambda}=1/\xi$. 

While $\xi$ could, on one hand, be determined by equating the analytical and cataloged coalescence energies for each event, we want to avoid the added obligation of uniquely defining $\xi$ for all GW events. On the other hand, we also wish to avoid an ad hoc parameter fitting scheme, such that $\xi$ is e.g. an averaged value of such calculations, particularly for events that are strongly PN-corrected due to e.g. tidal deformability or eccentricity. Therefore, we opt for a single representative calibration, maintaining that $\xi\in\mathbb{N}$. Our fiducial example is GW150914, which is ideal because there are no significant recorded eccentricies or tidal deformabilities. This yields a ``baseline" quadrupolar energy, upon which PN-correcting terms may be introduced if neccessary. From this singular ``analytical = catalog" matching, we obtain $\lfloor\xi\rfloor=3$ ($\lfloor\bullet\rfloor$ denotes the floor function), which we adopt for all subsequent comparisons. This is the source of the surrogate $R_\mathrm{eff}=-\sqrt{K}/6$, which modifies Eq. (\ref{endens}) as follows:
\begin{equation}\label{endens2}
T_{00}\simeq \frac{1}{8\pi G}\left(R_{00} +\frac{1}{12}g_{00}\sqrt{K_\mathrm{Kerr}} \right)=:\frac{1}{8\pi G}G^\mathrm{eff}_{00},
\end{equation}
where $R_{00}$ is given by Eq. (\ref{r00}), and $g_{00}$ is given in the Kerr metric.

\section{Results}

\subsection{The Approximate Energy}

Repeating the procedure in deriving Eq. (\ref{goo}), the effective $(00)$-th Einstein tensor component $G^\mathrm{eff}_{00}$ on the right-hand side of Eq. (\ref{endens2}), after implementing $K_\mathrm{Kerr}$ via Eq. (\ref{kretsch}) and $g_{00}$ in Eq. (\ref{kermet}), is Taylor expanded in small $a$ -- truncated at quadratic order to capture a leading spin correction -- and then integrated over $\theta$, readily removing the 1PM term:
\begin{equation}\label{goo2}
G_{00}^\mathrm{eff}\simeq \frac{2\pi G^2\mu^2}{r^4}\left(1.577-8.320 \frac{a^2}{r^2}\right).
\end{equation}
Substituting Eq. (\ref{goo2}) into Eq. (\ref{endens2}), we obtain the energy $E=T_{00}V$ at the shell radius $r=\rho$. We further define the surface energy at the time of merger $t=t_C$, for which $\rho=2GM$ and $\beta=\beta_C$:
\begin{equation}\label{energy}
\Rightarrow\quad E(t_C)\simeq 0.826\frac{\mu^2}{M}\left(1-5.276 \beta_C^2 \right).
\end{equation}

Comparing Eqs. (\ref{esimp}) and (\ref{energy}), the latter presents a systematic enhancement to the energy computation, yielding the anticipated GW150914 and GW170817 energy values of, respectively, $E(t_C)\simeq3.26 M_\odot $ and $E(t_C)\simeq0.14 M_\odot $. Furthermore, one may determine the maximum coalescence rotational velocity permitted by this approximation, in order to avoid unphysically negative energy values. The energy given by Eq. (\ref{energy}) vanishes at $\beta_C^2\simeq0.190$, which sets an upper bound, ceiling value on the normalized velocity of $\beta_C\leq0.435$. This is comparable to the rotational velocity at the total mass ISCO radius: $\beta_\mathrm{ISCO}=0.408$, which may serve as an extra constraint.

\subsection{Comparison with Select GW Signals}\label{sign}

The next objective is to test the precision and universality of Eq. (\ref{energy}). To this end, the model's approximate energy values for select, well-defined GW events are compared with the respective measured values, either directly as $E_\mathrm{rad}$ (as provided in GWTC-1 \cite{LIGOScientific:2018mvr}) or extracted from the difference between the total and final masses (i.e., the total-minus-remnant difference) $M-M_\mathrm{f}$ (c.f. $M$ and $M_\mathrm{f}$ in GTWC-2.1 through -4.0 \cite{LIGOScientific:2021usb, KAGRA:2021vkt, LIGOScientific:2025slb}). 

In total, 45 out of 368 cataloged events are to be analyzed: 45 representative events across GTWC-1 through -4.0, among these are noteworthy discoveries across the decade of GW astrophysics, and many were selected for having discrenible chirp rise trends in the strain plots in the \textit{GWOSC} open access catalog \cite{GWOSC}. One may suspect that this selection is inherently biased, and rightly so; it is therefore encouraged to pursue an extended analysis for \textit{all} cataloged events, even those with illegible strain plots with no visible chirp rise trends. However, for this introductory comparison between the mass-shell representation and previous observations, noteworthy and a sample of well-legible GW events serve collectively as a benchmark. 

Although the total and final masses have associating uncertainty ranges in the \textit{GWOSC} open access catalog \cite{GWOSC} and the GWTC papers \cite{LIGOScientific:2018mvr, LIGOScientific:2021usb, KAGRA:2021vkt, LIGOScientific:2025slb}, the total-minus-remnant difference will be calculated only through the central mass values (i.e., naively ignoring their uncertainties). In addition, the approximate values are determined using the central (best-fit) values of the CCB masses $m_1$ and $m_2$, together with the anticipated value of the normalized rotational speed at merger: $\beta_C=GMf_\mathrm{GW,peak}$. The peak frequency $f_\mathrm{GW,peak}$ is either reported in the relevant detection papers or inferred from the strain plots in the \textit{GWOSC} open access catalog  (see Section \ref{ill} for approaching illegible strain plots in this analysis). 

Agreement between the approximated energy via the mass-shell representation and the respective energy value inferred from observation is quantified by computing a 1:1 ratio of the central energy values: $\mathrm{Ratio}:=\mathrm{(smaller~value)/(larger~value)}\in(0,1)$. Here, $\mathrm{Ratio}\rightarrow0$ indicates no alignment between the values, whereas $\mathrm{Ratio}\rightarrow1$ indicates (near-)perfect agreement. However, it is essential to complement these ratios with further goodness-of-fit, statistical tests. These are to be discussed in Section \ref{goods}.

Tables \ref{t1} and \ref{t2} contain all cataloged events listed in the GWTC-1 and -2.1 papers \cite{LIGOScientific:2018mvr, LIGOScientific:2021usb}, including GW190521 \cite{LIGOScientific:2020iuh} and GW190814 \cite{LIGOScientific:2020zkf}. Table \ref{t3} presents 11 select events from the GWTC-3 catalog \cite{KAGRA:2021vkt}, while Table \ref{t4} shows 9 select events from the GWTC-4.0 catalog \cite{LIGOScientific:2025slb} along with 4 noteworthy O4 discoveries \cite{LIGOScientific:2024elc, LIGOScientific:2025rsn, LIGOScientific:2025brd, LIGOScientific:2025rid}. All tables list the CCB masses, the measured/extracted radiated energy, the analytical prediction via Eq. (\ref{energy}), and the corresponding 1:1 ratio. 

\begin{table*}%[h!]
\caption{\label{t1} All cataloged events in Table III of the GWTC-1 paper \cite{LIGOScientific:2018mvr}, which consist of detected GW signals from the first and second observation runs. The peak frequencies can be found in the detection papers as well as from the strain plots provided in the open-access catalog \cite{GWOSC}, and the mass and energy parameters are sourced from Ref. \cite{LIGOScientific:2018mvr}. The approximate energy via Eq. (\ref{energy}) utilized the center and estimated values, with a corresponding 1:1 ratio with the measured central energy value.}
%\begin{ruledtabular}
\begin{tabular}{cccccc}
 &\multicolumn{3}{c}{c.f. GWTC-1 \cite{LIGOScientific:2018mvr} } &     \\
 Event & $m_1\,(M_\odot)$ & $m_2\,(M_\odot)$  & $E_\mathrm{rad}/(M_\odot )$ &  Eq. (\ref{energy})$/(M_\odot )$ & $1:1$ Ratio\\ \hline
 GW150914 \cite{LIGOScientific:2016aoc} & $35.6^{+4.7}_{-3.1}$   &   $30.6^{+3.0}_{-4.4}$   & $3.1^{+0.4}_{-0.4}$  & $3.26$ & $0.951$\\
 LVT151012 & $23.2^{+14.9}_{-5.5}$ &  $13.6^{+4.1}_{-4.8}$  &  $1.6^{+0.6}_{-0.5}$ &   $1.65$   & $0.970$ \\
 GW151226 \cite{LIGOScientific:2016sjg} & $13.7^{+8.8}_{-3.2}$   &  $7.7^{+2.2}_{-2.5}$   &    $1.0^{+0.1}_{-0.2}$ &  $0.927$ & $0.927$\\
 GW170104 \cite{LIGOScientific:2017bnn} & $30.8^{+7.3}_{-5.6}$   & $20.0^{+4.9}_{-4.6}$	  &   $2.2^{+0.5}_{-0.5}$   & $2.29$ & $0.959$\\
 GW170608 \cite{LIGOScientific:2017vox} & $11.0^{+5.5}_{-1.7}$   & $7.6^{+1.4}_{-2.2}$	 &    $0.9^{+0.0}_{-0.1}$  & $0.895$ & $0.995$   \\
 GW170729 & $50.2^{+16.2}_{-10.2}$ & $34.0^{+9.1}_{-10.1}$  & $4.8^{+1.7}_{-1.7}$	&    $3.97$ & $0.828$\\
 GW170809 & $35.0^{+8.3}_{-5.9}$     & $23.8^{+5.1}_{-5.2}$  & $2.7^{+0.6}_{-0.6}$	&   $2.75$ & $0.980$\\
 GW170814 \cite{LIGOScientific:2017ycc} & $30.6^{+5.6}_{-3.0}$     & $25.2^{+2.8}_{-4.0}$  & $2.7^{+0.4}_{-0.3}$	& $2.76$  & $0.980$\\
 GW170817 \cite{LIGOScientific:2017vwq} & $1.46^{+0.12}_{-0.10}$ & $1.27^{+0.09}_{-0.09}$  & $\geq0.04$	  & $0.140$ & - -\\
 GW170818 & $35.4^{+7.5}_{-4.7}$     & $26.7^{+4.3}_{-5.2}$ &  $2.7^{+0.5}_{-0.5}$	 & $2.99$  & $0.905$\\
 GW170823 & $39.5^{+11.2}_{-6.7}$   & $29.0^{+6.7}_{-7.8}$ &  $3.3^{+1.0}_{-0.9}$	 & $3.29$ & $0.997$\\
\end{tabular}
%\end{ruledtabular}
\end{table*}

\begin{table*}
\caption{\label{t2} All cataloged events listed in Table VI of the GWTC-2.1 paper \cite{LIGOScientific:2021usb} (with GW190521 \cite{LIGOScientific:2020iuh} and GW190814 \cite{LIGOScientific:2020zkf}), which consist of detected GW signals from the first half of the third observation run. The peak frequencies are mainly estimates from the open-access strain plots provided in Ref. \cite{GWOSC}, and the mass parameters are sourced from Ref. \cite{LIGOScientific:2021usb}. The surrogate value for the radiated GW energy follows the mass difference between the total and final masses. The approximate energy via Eq. (\ref{energy}) utilized the center and estimated values, with a corresponding 1:1 ratio with the calculated mass difference.}
%\begin{ruledtabular}
\begin{tabular}{cccccc}
 &\multicolumn{3}{c}{c.f. GWTC-2.1 \cite{LIGOScientific:2021usb} }    \\
 Event & $m_1\,(M_\odot)$ & $m_2\,(M_\odot)$  & $M-M_\mathrm{f}\,(M_\odot)$ &  Eq. (\ref{energy})$/(M_\odot )$ & $1:1$ Ratio\\ \hline
 GW190403\_051519 & $85.0^{+27.8}_{-33.0}$   &   $20.0^{+26.3}_{-8.4}$   & $4.40$  & $2.05$  & $0.466$  \\
 GW190426\_190642 & $105.5^{+45.3}_{-24.1}$ &  $76.0^{+26.2}_{-36.5}$  & $9.40$  &   $8.72$  & $0.928$ \\
 GW190521 \cite{LIGOScientific:2020iuh} & $98.4^{+33.6}_{-21.7}$ & $57.2^{+27.1}_{-30.1}$ & $5.70$ & $6.73$ & $0.847$	\\
 GW190725\_174728 & $11.8^{+10.1}_{-3.0}$   &  $6.3^{+2.1}_{-2.5}$   &  $0.70$   & $0.769$ & $0.911$ \\
 GW190805\_211137 & $46.2^{+15.4}_{-11.2}$   & $30.6^{+11.8}_{-11.3}$   & $4.30$   & $3.58$ & $0.833$ \\
 GW190814 \cite{LIGOScientific:2020zkf} & $23.2^{+1.1}_{-1.0}$   & $2.59^{+0.08}_{-0.09}$ & $0.20$   & $0.174$ & $0.868$    \\
 GW190916\_200658 & $43.8^{+19.9}_{-12.6}$   & $23.3^{+12.5}_{-10.0}$ & $3.00$  & $2.79$  & $0.931$   \\
 GW190917\_114630 & $9.7^{+3.4}_{-3.9}$ & $2.1^{+1.1}_{-0.4}$  &  $0.20$  & $0.209$  & $0.959$ \\
 GW190925\_232845 & $20.8^{+6.5}_{-2.9}$   & $15.5^{+2.5}_{-3.6}$  & $1.80$   & $1.79$  & $0.996$ \\
 GW190926\_050336 & $41.1^{+20.8}_{-12.5}$   & $20.4^{+11.4}_{-8.2}$  & $2.30$  & $2.49$ & $0.923$ \\
\end{tabular}
%\end{ruledtabular}
\end{table*}

\begin{table*}
\caption{\label{t3} 11 select GW events out of 36 cataloged events in Table IV of the GWTC-3 paper \cite{KAGRA:2021vkt}, which consist of detected GW signals from the second half of the third observation run. These events were mainly selected for their pronounced strain signals provided in the open-access catalog \cite{GWOSC}. The peak frequencies are mainly estimates, and the mass and energy parameters are sourced from Ref. \cite{KAGRA:2021vkt}. The surrogate value for the radiated GW energy follows the mass difference between the total and final masses. The approximate energy via Eq. (\ref{energy}) utilized the center and estimated values, with a corresponding 1:1 ratio with the calculated mass difference.}
%\begin{ruledtabular}
\begin{tabular}{cccccc}
 &\multicolumn{3}{c}{c.f. GWTC-3 \cite{KAGRA:2021vkt} }   \\
 Event & $m_1\,(M_\odot)$ & $m_2\,(M_\odot)$  & $M-M_\mathrm{f}\,(M_\odot)$  & Eq. (\ref{energy})$/(M_\odot )$ & $1:1$ Ratio\\ \hline
 GW191204\_171526 &  $11.7^{3.3}_{-1.7}$  & $8.4^{+1.3}_{-1.7}$ & $1.01$ & $0.980$  & $0.970$    \\
 GW191215\_223052 & $24.9^{+7.1}_{-4.1}$ & $18.1^{+3.8}_{-4.1}$ & $1.90$    & $2.09$ & $0.909$   \\
 GW191216\_213338 &  $12.1^{+4.6}_{-2.2}$  & $7.7^{+1.6}_{-1.9}$ & $0.93$     & $0.922$ & $0.991$ \\
 GW191222\_033537 & $45.1^{+10.9}_{-8.0}$ & $34.7^{+9.3}_{-10.5}$ & $3.50$ & $3.88$  & $0.902$   \\
 GW191230\_180458 & $49.4^{+14.0}_{-9.6}$ & $37.0^{+11.0}_{-12.0}$ & $4.00$ & $4.24$   & $0.943$    \\
 GW200112\_155838  & $35.6^{+6.7}_{-4.5}$ & $28.3^{+4.4}_{-5.9}$ & $3.10$ & $3.06$ & $0.987$  \\
 GW200129\_065458 & $34.5^{+9.9}_{-3.1}$ & $29.0^{+3.3}_{-9.3}$ & $3.10$  &  $3.08$  & $0.993$  \\
 GW200224\_222234 & $40.0^{+6.7}_{-4.5}$ & $32.7^{+4.8}_{-7.2}$ & $3.60$  & $3.52$ & $0.978$  \\ 
 GW200225\_060421 & $19.3^{+5.0}_{-3.0}$ & $14.0^{+2.8}_{-3.5}$ & $1.40$  & $1.61$  & $0.869$  \\
 GW200302\_015811 & $37.8^{+8.7}_{-8.5}$ & $20.0^{+8.1}_{-5.7}$ & $2.30$  &  $2.39$  & $0.962$  \\
 GW200311\_115853 & $34.2^{+6.4}_{-3.8}$ & $27.7^{+4.1}_{-5.9}$ & $2.90$  & $3.04$  & $0.955$  \\
\end{tabular}
%\end{ruledtabular}
\end{table*}

\begin{table*}
\caption{\label{t4} 13 select GW events out of the 129 cataloged O4 events, listed in Table III of the GWTC-4.0 paper \cite{LIGOScientific:2025slb} (which consist of detected GW signals from the first half of the fourth observation run) and noteworthy O4 discoveries. These events were mainly selected for their pronounced strain signals provided in the open-access catalog \cite{GWOSC}. The peak frequencies are mainly estimates, and the mass and energy parameters are sourced from Ref. \cite{LIGOScientific:2025slb}. The surrogate value for the radiated GW energy follows the mass difference between the total and final masses. The approximate energy via Eq. (\ref{energy}) utilized the center and estimated values, with a corresponding 1:1 ratio with the calculated mass difference.}
%\begin{ruledtabular}
\begin{tabular}{cccccc}
 &\multicolumn{3}{c}{c.f. GWTC-4.0 \cite{LIGOScientific:2025slb} }   \\
 Event & $m_1\,(M_\odot)$ & $m_2\,(M_\odot)$  & $M-M_\mathrm{f}\,(M_\odot)$  & Eq. (\ref{energy})$/(M_\odot )$ & $1:1$ Ratio \\ \hline
 GW230814 \cite{LIGOScientific:2024elc} & $33.6^{+2.8}_{-2.2}$    & $28.3^{+2.1}_{-3.0}$  & $2.90$  & $2.83$  & $0.976$   \\
 GW230924\_124453 & $28.8^{+5.9}_{-4.0}$ & $23.1^{+4.4}_{-4.4}$  &  $2.50$  &  $2.59$ & $0.965$    \\
 GW231008\_142521 &  $45^{+17}_{-12}$   & $25.5^{+10.7}_{-9.7}$  &  $3.00$     & $3.07$  & $0.977$  \\
 GW231123 \cite{LIGOScientific:2025rsn} & $137^{+23}_{-18}$  & $101^{+22}_{-51}$ &  $14.0$ &  $10.9$  &  $0.779$   \\
 GW231226\_101520 & $40.1^{+4.4}_{-2.9}$  & $35.0^{+3.2}_{-4.9}$  & $3.40$  &  $3.50$   &   $0.971$   \\
 GW231230\_170116  & $54^{+50}_{-15}$  & $35^{+15}_{-14}$  & $3.00$  & $4.16$  & $0.721$   \\
 GW231231\_154016 & $22.5^{+5.7}_{-3.3}$ & $17.2^{+2.9}_{-3.2}$  &  $1.80$  &  $1.96$   & $0.918$   \\
 GW240104\_164932 & $42.3^{+9.4}_{-6.7}$  & $32.1^{+7.5}_{-8.0}$  & $3.70$   & $3.53$  & $0.954$   \\ 
 GW240107\_013215 & $59^{+27}_{-18}$  & $32^{+20}_{-16}$  &  $5.00$  & $3.81$  &  $0.762$  \\
 GW240109\_050431 & $28.8^{+7.5}_{-6.2}$  & $18.1^{+4.8}_{-4.1}$  & $2.00$  & $2.14$   & $0.935$   \\
 GW241001 \cite{LIGOScientific:2025brd}  & $19.1^{+3.6}_{-2.5}$  & $5.9^{+0.8}_{-0.8}$  & - -   & $0.671$   &  - - \\
 GW241110 \cite{LIGOScientific:2025brd} & $17.2^{+5.0}_{-4.4}$  & $7.7^{+2.2}_{-1.5}$  &  - -   &  $0.935$  &  - -  \\
 GW250114 \cite{LIGOScientific:2025rid} & $33.5^{+1.2}_{-0.8} $ & $32.2^{+0.9}_{-1.3}$  &  $3.10$  & $3.29$  &  $0.942$  \\
\end{tabular}
%\end{ruledtabular}
\end{table*}

\subsubsection{On Events with Illegible Strain Plots}\label{ill}

For events in which $f_\mathrm{GW,peak}$ is neither reported in the detection paper nor cleanly discernible from the chirp rise in the strain plots (i.e., LVT151012 from GWTC-1 and in Table \ref{t1}, and GW190403\_051519 and GW190917\_114630 from GWTC-2.1 and in Table \ref{t2}) -- while having mass measurements --, only the leading Newtonian contribution in Eq. (\ref{energy}) is calculated. As it is known, this is an incomplete calculation, but it is proven for least-massive systems that it suffices. 

Consider the events GW190521 (sourced by BBHs, in Table \ref{t2}) and GW170817 (sourced by BNSs, in Table \ref{t1}); both have discernible strain plots. For these two events, we calculate their respective rotational speed ratios at merger $\beta_C$ and its square. For GW190521, $\beta_C=0.0769$, obtained using $f_\mathrm{GW,peak}\simeq100$ Hz from the \textit{GWOSC} open access catalog; this calculates $\beta_C^2=0.00591$. For GW170817, $\beta_C=0.00405$, obtained from $f_\mathrm{GW,peak}\simeq300$ Hz, calculating $\beta_C^2=1.64\times10^{-5}$. Even though the speed ratio-squared is scaled by $\sim5.3$ in the quadratic spin contribution of Eq. (\ref{energy}), the quadratic contribution remains to be a small factor, essentially making the Newtonian contribution the leading factor in the energy calculation. Thus, $1-5.3\beta_C^2\approx1$ for these two examples, which enables one to use (within reason, see footnote\footnote{Of course, one must recognize that $\beta_C\propto M$, and the total mass strengthens the role of the quadratic spin contribution. Therefore, one may argue that the spin contribution is miniscule for least-massive systems and more noticable for more-massive systems.}) the leading Newtonian contribution for cataloged events with illegible strain plots. 

Only three of the 45 representative events have illegible strain plots on the \textit{GWOSC} open access catalog: LVT151012 from GWTC-1, and GW190403\_051519 and GW190917\_114630 from GWTC-2.1. The approximated coalescence energy for these events are calculated only from the leading Newtonian contribution in Eq. (\ref{energy}), on the basis that the analysis above holds for these three events. However, it is ill-advised to assume, effectively, $\beta_C^2\ll1$ even for events that have substantial mass measurements yet have illegible strain plots; the reason being $\beta_C\propto M$, and larger mass systems would have a noteable spin contribution in the coalescence energy. This would lead to calculations where the spin information is not accounted for, and thus not reflected in the approximated coalescence energy if naively omitted. 

This limitation is the reason that only 11 GWTC-3 events were included in Table \ref{t3} (and 13 in Table \ref{t4}, although these are representative of the 129 cataloged events overall that may or may not have discernible strain plots); the remaining events exhibit illegible, if not indeterminable, strain plots in the \textit{GWOSC} open access catalog. Therefore, it is encouraged to derive a ``rule-of-thumb" peak frequency expression from first principles, to use in cases where the strain plots are not cleanly legible. This is beyond the scope of the present work, but deferred for a future study.

\section{Discussion}\label{disc}

The analytical effort in this work reinterprets the effective one-body approach, which traditionally treats coalescence as the inspiraling plunge path of the reduced mass particle (analogous to a marble in a funnel), by instead considering a rotating and shrinking shell with a smeared mass distribution of constant reduced mass measure. The focus is placed on the energy density of the CCB mass shell, whose derivation employs methods akin to quantum-mechanical variational principle. Using a Kerr metric Ansatz and an effective Ricci scalar defined through a scaled, negated square-root of the Kerr-metric Kretschmann scalar, the approximation for the coalescence energy (Eq. [\ref{energy}]) was applied to a set of cataloged GW signals to test its precision and generality. Tables \ref{t1}--\ref{t4} show this agreement between the approximation and observation via a set of 1:1 ratios from the central energy values.

The 1:1 ratios span the range $0.828\sim0.997$ for 38 events, indicating generally strong agreement. Exceptions include three events with ratios in the range $0.721\sim0.779$ (all from the O4 run; see Table \ref{t4}), three events for which a ratio could not be determined due to either an unconstrained radiated energy (GW170817, see Table \ref{t1}) or an unknown total-minus-remant mass difference (GW241001 and GW241110, see Table \ref{t4}), and GW190403\_051519, which yields a ratio of 0.462 (see Table \ref{t2}). In the case of the three O4 events with ratios in the $0.721\sim0.779$ range, the associated uncertainties from observation are relatively large, and the observational central values exceed the approximated in all but GW231230\_170116 (see footnote\footnote{When the approximated value exceeds the observational value, any visible shift in energy may reasonably be attributed to PN corrections. When, instead, the observational value exceeds the approximation, the full uncertainty ranges in the component masses and in the total and final masses -- found from observation -- must be carefully considered.}).

In the case of GW190403\_051519, the strain plots in the open access catalog show no legible chirp-rise structure in either detector, and the corresponding mass measurements from observation carry relatively large uncertainty ranges. Notably, the observed energy output exceeds the approximate value by nearly a factor of 2 -- well beyond a discrepancy that could be reasonably be attributed to the approximate nature of the model or to a simple PN-corrected energy shift. In general, large measurement uncertainties indicate that the quantity in question is poorly constrained, which can naturally give rise to such ambiguities. For these reasons, it is not unreasonable to treat the 1:1 ratio associated with GW190403\_051519 as an outlier.

Nevertheless, Eq. (\ref{energy}) clearly captures only part of the full coalescence-energy picture. Among the selected GW events are systems known to exhibit non-zero eccentricity (e.g. GW190521, see Table \ref{t2}) or measurable tidal deformability (e.g. GW170817, see Table \ref{t1}). It is therefore reasonable to infer that the residual discrepancies between approximation and observation motivate the inclusion of systematic, PN-inspired correction terms in Eq. (\ref{energy}) (see footnote\footnote{This applies to GW events with 1:1 ratios below $0.955$, the median of the 38 ``agreeing" ratios.}).

\subsection{Goodness-of-Fit Between Mass-Shell and Observation}\label{goods}

While the 1:1 ratios give us a preliminary sense of agreement between the mass-shell representation and observation, we must consider statistical goodness-of-fit tests to help quantify the level of agreement. Of the 45 representative events analyzed in this comparison, 42 have well-defined 1:1 ratios in Tables \ref{t1}--\ref{t4} regardless of its closeness to unity. It is these events that are taken into account in this goodness-of-fit, statistical study.

Goodness-of-fit tests between central predicted measurements $P_i$, i.e. $E(t_C)$ via Eq. (\ref{energy}), and their central observed counterparts $O_i$, i.e. $E_\mathrm{rad}$ or $M-M_\mathrm{f}$, over $n$ events include, yet are not limited to:
\begin{itemize}
\numparts
\item Root Mean Squared Error (RMSE), having the same units as the data:
\begin{equation}
\mathrm{RMSE}=\sqrt{\frac{1}{n}\sum_{i=1}^n(P_i-O_i)^2},
\end{equation}
\item Mean Absolute Error (MAE), robust to outliers and also having the same units as the data:
\begin{equation}
\mathrm{MAE}=\frac{1}{n}\sum_{i=1}^n \left|P_i-O_i \right|,
\end{equation}
\item Coefficient of Determination $R^2$:
\begin{equation}
R^2=1-\frac{\sum (O_i-P_i)^2}{\sum(O_i-\bar{O})^2}
\end{equation}
($\bar{O}$ is the average value of all observed values), 
\item Mean Absolute Percentage Error (MAPE):
\begin{equation}
\mathrm{MAPE}=\frac{100}{n}\sum_{i=1}^n \left|\frac{P_i-O_i}{O_i} \right|,
\end{equation}
and finally 
\item The $\chi^2$ statistic for central measurements (and its reduced form over the degrees of freedom $\mathrm{df}=n-1$):
\begin{equation}
\chi^2=\sum_{i=1}^n\frac{(O_i-P_i)^2}{P_i},\quad\chi^2_\mathrm{red}=\frac{1}{n-1}\chi^2.
\end{equation}
\endnumparts
\end{itemize}
We consider all of the above to better quantify the mass-shell model's agreement with observation. We furthermore consider a second type of ratio: the predicted-to-observed ratio $P_i/O_i$, discerning how far above or below unity individual ratios are. We introduce this secondary ratio in order to compute the mean and median $P_i/O_i$ ratios as yet another insurance test for goodness-of-fit.

\begin{table*}
\centering
\caption{\label{tab:goods} Statistical goodness-of-fit tests on 42 GW events from Tables \ref{t1}--\ref{t4} with well-defined 1:1 ratios and their computed values.}
%\begin{ruledtabular}
\begin{tabular}{c||c}
  Test & Value \\ \hline
% -- - - - - - - - - 
  RMSE &  $0.7114M_\odot$\\ 
  MAE & $0.3517M_\odot$ \\ 
  $R^2$ & 0.9083 \\ 
  MAPE & $8.86\%$ \\ 
  $\chi^2$ & 5.0132 \\ 
  $\chi^2_\mathrm{red}$ & 0.1223 \\ 
  Mean $P_i/O_i$ & 1.0009 \\ 
  Median $P_i/O_i$ & 1.0228 
\end{tabular}
%\end{ruledtabular}
\end{table*}

\begin{figure*}
\centering
	\includegraphics[width=0.8\textwidth]{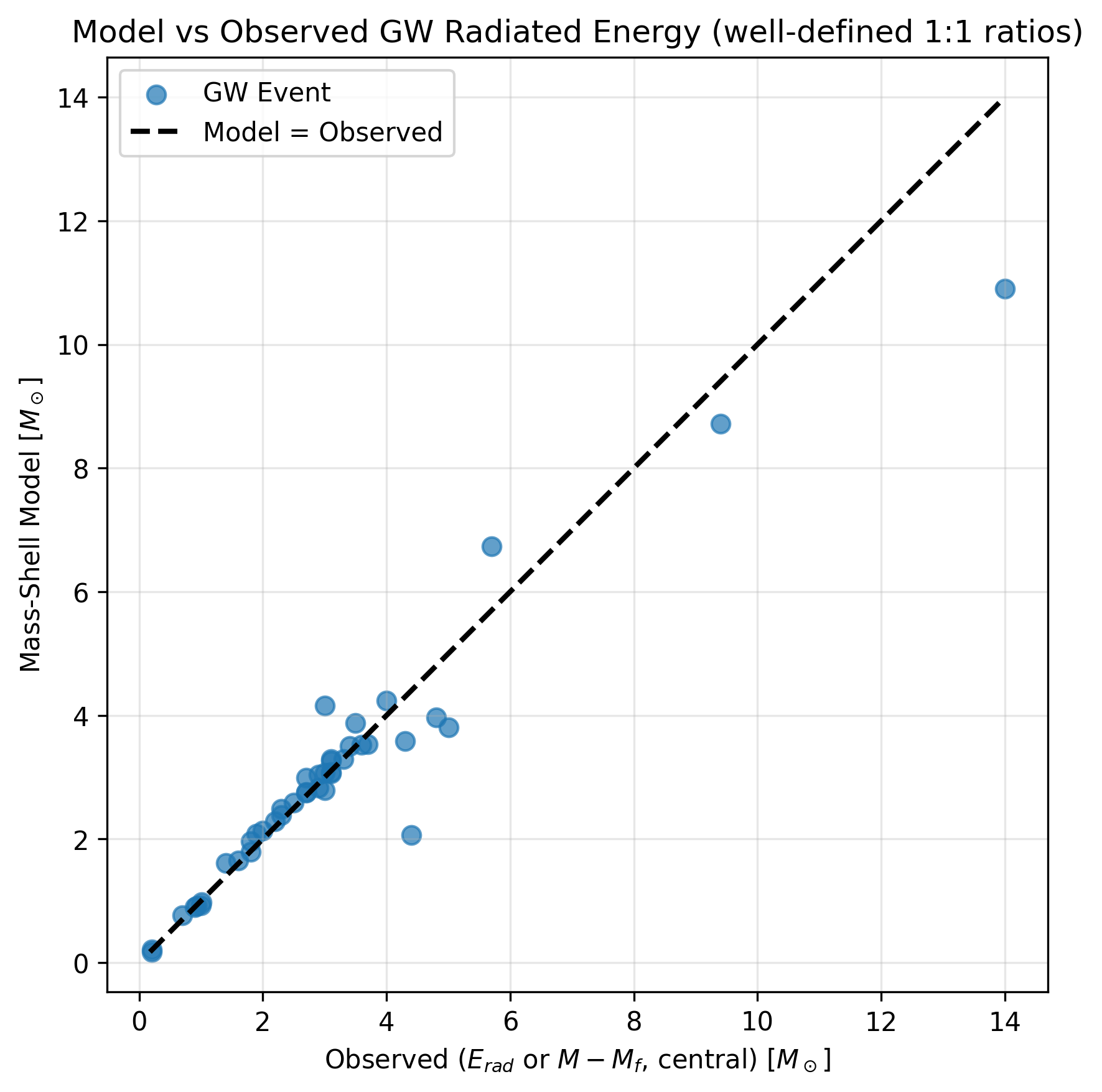}
\caption{\label{fig:scats} Scatter plot of 42 GW events (blue dots) from Tables \ref{t1}--\ref{t4} with well-defined 1:1 ratios; x-axis corresponds to the observed energy (central value), and the y-axis corresponds to the model prediction. Black-dashed line is the $y=x$ line for perfect mutuality between model and observation.}
\end{figure*}

Table \ref{tab:goods} provides the metrics of each statistical goodness-of-fit test between the mass-shell representation and observation across 42 of the 45 representative GW events, and Figure \ref{fig:scats} provides a scatter plot for each respective event with corresponding model-predicted and observed energy values. Given the metrics provided in Table \ref{tab:goods}, one can claim that mass-shell model demonstrates an excellent quantitative agreement with the observed radiated energies.

 Using central values only, the model achieves a high coefficient of determination ($R^2=0.9083$), low absolute and relative errors ($\mathrm{MAPE}=8.86\%$, $\mathrm{RMSE}=0.7114M_\odot$ and $\mathrm{MAE}=0.3517M_\odot$) given the (well constrained) observed radiated energies range from $\in[0.20,\,14.0]M_\odot$, and near-unity median and mean predicted-to-ratios (mean: $1.0009$, median: $1.0228$). The $\chi^2$ statistic of 5.0132 (reduced $\chi^2_\mathrm{red}=0.1223$) further confirms that residuals are small relative to the model's calculations. This close agreement is visually apparent in Figure \ref{fig:scats}, where one sees how each blue point (each event) clusters near (or deviates from) the black-dashed 1:1 (``model = observed") line. 

Collectively, these metrics indicate that the model captures the underlying physics of energy radiation in compact binary mergers to high accuracy, readily at the provided leading order, with only modest scatter and essentially no net bias.

\subsection{Collision-Channel Mass/Energy Conservation}\label{coll}

Viewing CCBs through a particle-physics lens is not new. For instance, the EOB framework and its associating Hamiltonian are inspired by quantum field theory (QFT) \cite{Buonanno:1998gg, Damour:2009zoi, Damour:2012mv}, with extensions to include e.g. tidal deformation \cite{Hinderer:2016eia, Steinhoff:2016rfi, Steinhoff:2021dsn} employing effective field theory (EFT) to describe tidally induced, PN-corrected phase shifts via a ``non-geodesic" mass-shell condition $p_\mu p^\mu-\mu^2=\mu^2_\mathrm{NG}$. This modification imposes an altered dispersion relation in which the shifted frequency manifests as tidal phase corrections. More recently, a worldline QFT approach \cite{Mogull:2020sak} and its subsequent developments, e.g. \cite{Driesse:2024feo}, attempt to describe the PN regime of coalescence using Feynman diagrams, calculating scattering amplitudes from which end-state measurables, such as relative scattering angle and inspiral-phase radiated energy, may be derived \cite{Driesse:2024feo}. Within this logic, the merger phase in IMR dynamics can be effectively described as either a $2\rightarrow1$ (i.e. $\mathrm{CCB}\rightarrow\mathrm{BH}$) inelastic collision channel or a $2\rightarrow2$ (i.e. $\mathrm{CCB}\rightarrow\mathrm{BH+radiation}$) elastic collision channel (see e.g. Ref. \cite{Aoki:2024boe}). 

For a source-frame, head-on collision at the merger phase, mass-energy conservation before and after the collision is expected to be elastic:
\begin{equation}
m_1+m_2\equiv M=E_\mathrm{rad}+M_\mathrm{f}.
\end{equation}
Recognizing that the ``parent" constituents $m_1$ and $m_2$ are converted into the ``offspring" constituents $M_\mathrm{f}$ and $E_\mathrm{rad}$, any energy radiated during CCB merger is simply the residual mass-energy associated with the mass difference $M-M_\mathrm{f}$. This principle is well understood, and this is precisely the total-minus-remnant difference used as the energy surrogate in Tables \ref{t2}--\ref{t4} for comparison with the approximation. However, as seen in these tables, certain events exhibit discrepancies between approximation and observation. Assuming that this is not merely a consequence of working with approximate values, particularly in cases where the approximate value exceeds the observed value, these apparent ``negative" energy shifts are most likely attributed to radiative PN corrections. With this in mind, suppose $E_\mathrm{rad}$ takes the form of $E(t_C)$ via Eq. (\ref{energy}) minus an energy shift $\Delta E$:
\begin{equation}\label{emc2}
 E_\mathrm{rad}\equiv M-M_\mathrm{f}=0.826\alpha \mu \left(1-5.276 \beta_C^2 \right)-\Delta E.
\end{equation}
This correction is naturally imprinted in the coalescence speed ratio. From the mass-shell model review in Section \ref{model}, the surface-level speed ratio emerges from the ratio between the Kerr spin parameter and the surface radius $a/\rho$, with $a=J/\mu$. Under $D\gg S$, the mass shell is effectively viewed as a point mass with angular momentum $J=I\Omega$ and moment of inertia $I=\mu\rho^2$. However, the angular frequency $\Omega$ should inherit PN corrections, for the angular frequency in the mass-shell picture is exactly the orbital frequency in the standard two-body coalescence picture, recall Figure \ref{fig:mod}. And from this orbital frequency, we have a PN-expanded orbital phase $\Psi$, written here following Refs. \cite{Favata:2013rwa, Flanagan:2007ix, Chatziioannou:2020pqz}: 
\begin{eqnarray}\label{phase}
\Psi=&~\Omega t_C+\varphi_C-\frac{\pi}{4}\\\nonumber
&+\frac{3}{128\alpha u^5}\left(\delta\Psi_\mathrm{2PN}^\mathrm{ecc.}+\delta\Psi_\mathrm{3PN}^\mathrm{spin}+\delta\Psi_\mathrm{3.5PN}^\mathrm{pp}+\delta\Psi_\mathrm{6PN}^\mathrm{tm}+\delta\Psi_\mathrm{6PN}^\mathrm{tidal} \right).
\end{eqnarray}
Here, $\varphi_C$ is the orbital phase at coalescence, and $u=(\Omega GM)^{1/3}$ is a dimensionless expansion variable such that the PN correction order $n$ is set by coefficients of $u^{2n}$; the $\pi/4$ contribution is omitted in Ref. \cite{Favata:2013rwa}. In increasing order of PN correction: the 2PN phase shift is due to eccentricity (e.g. Eq. (3) in Ref. \cite{Favata:2013rwa}), the 3PN term due to axial spin effects (e.g. Eq. (2) in Ref. \cite{Favata:2013rwa}), the 3.5PN term due to $\alpha$-dependent point-particle (pp) contributions $1+\sum_{n=2}^7 c_n(\alpha)u^n$, and the two 6PN terms respectively due to $\alpha$-independent test-mass (tm) limitations $\sum_{n=8}^{12}c_n u^n$ and due to tidal effects.

On the other hand, another mapping scheme for characterizing the energetics of CCBs, besides the EOB Hamiltonian, is the CCB Lagrangian (e.g. \cite{Abdelsalhin:2018reg}), which governs the motion of the CCB:
\begin{equation}
\mathcal{L}(z^i,\dot{z}^i,\ddot{z}^i,M_A, J_A, Q^L_A, \dot{Q}^L_A, S^L_A, \dot{S}^L_A).
\end{equation}
Here, $z^i=z_1^i-z_2^i$ is the relative position of the binary components (the index $A=1,2$ denotes the individual components), $M_A,~J_A$ are respectively the mass and current of object $A$, and $Q^L_A,~S_A^L$ are respectively the $L$-th rank Thorne multipole mass and current moments of object $A$. The full Lagrangian (in shorthand $\mathcal{L}$) is summed over $A$ for the orbital motion of the compact binary $\mathcal{L}_\mathrm{orb}$; if one or both component masses are NSs an additional internal Lagrangian $\mathcal{L}_2^\mathrm{int}$ is included to account for the tidal response induced by the companion.  While the Lagrangian is classically the difference between the kinetic and gravitational potential energies, the conserved energy in the system $E$ is its direct sum (i.e., one changes the sign of all potential energy terms in $\mathcal{L}$ to express $E$). Ref. \cite{Hinderer:2009ca} writes this energy in two explicit forms (their Eq. (17) and Eq. (A.3)), both highlighting the tidal contributions to the PN expansion. These can be combined into a single linearly expanded form, with $G=1$, as:
\begin{eqnarray}\label{enpn}
E=-\frac{1}{2}\alpha M^{5/3}\Omega^{2/3}\Big[1&-\frac{9\lambda_1 m_2\Omega^{10/3}}{M^{5/3}m_1}\left(1+5\frac{\Omega^2}{\Omega_0^2} \right)+1\leftrightarrow2\\\nonumber
&+\left(\mathrm{PN~pp~corrections} \right)\\\nonumber
&-\left(\mathrm{octopolar~tides}\right)\\\nonumber
&-\left(\mathrm{nonlinear~hydrodynamics}\right) \Big].
\end{eqnarray}
The overall negative sign encodes the fact that the energy is radiating away from the CCB. In this expression, $1\leftrightarrow2$ denotes the exchange of integer indices for the dimensionful tidal deformability terms containing $\lambda_i:=\Lambda_i\cdot(m_i)^5$ and the CCB masses, and $\Omega_0$ is the quadrupolar $f$-mode angular frequency. Additional PN corrections are left implicit for brevity.

Note that $\alpha^{3/5}M$ is the chirp mass $\mathcal{M}$, such that $E\propto\mathcal{M}^{5/3}$. To the untrained eye, there exists a curious, ad hoc correlation between the radiated energy and one-tenth of the chirp mass $\mathcal{M}$ (see e.g. \cite{GWOSC} and the comparison between $M-M_\mathrm{f}$ and $\sim\mathcal{M}/10$ listed in Refs. \cite{LIGOScientific:2018mvr, LIGOScientific:2021usb, KAGRA:2021vkt, LIGOScientific:2025slb}). Although Ref. \cite{MacKay:2024qxj} demonstrated that this correlation is not systematically reliable -- and this is reinforced in the present work via Eqs. (\ref{esimp}) and (\ref{energy}) --, Eq. (\ref{enpn}) nonetheless confirms that the correlation between the radiated energy and the chirp mass is not entirely ad hoc. Rather, it shows that a direct proportionality $E\sim\mathcal{M}$ is not physically justified. 

However, when expressed in terms of the dimensionless PN variable $u=(\Omega GM)^{1/3}$, recovering $G$, and defining $x=u^2$ (so that the $n$PN correction corresponds to a term of order $x^n$), the chirp mass vanishes in lieu of a direct proportionality between the energy and the CCB's \textit{total} mass. Thus, Eq. (\ref{enpn}) can be rewritten in a form strucurally analogous to the linear expansion in Eq. (\ref{phase}), allowing us to identify the individual PN contributions to the inspiral energy:
\begin{eqnarray}\label{enpn2}
\Rightarrow\quad E=-\frac{1}{2}\alpha Mx\Big[&1+\frac{1}{x}\Big(\Delta_\mathrm{2PN}^\mathrm{ecc.}+\Delta_\mathrm{3PN}^\mathrm{spin}\\\nonumber
&+\Delta_\mathrm{3.5PN}^\mathrm{pp}+\Delta_\mathrm{6PN}^\mathrm{tm}+\Delta_\mathrm{6PN}^\mathrm{tidal} \Big)\Big].
\end{eqnarray}
Comparing Eq. (\ref{enpn2}) to Eq. (\ref{emc2}), we see clear structural parallels: all expressions exhibit a direct dependence on the symmetric mass ratio $\alpha$, all are expressed as a mass-energy relation (Eq. (\ref{emc2}) via the reduced mass and Eq. (\ref{enpn2}) via the total mass), and all are trailed by subtle energy contributions sourced by various yet relevant PN corrections. We also note that Eq. (\ref{enpn2}) is explicitly frequency-dependent through the pre factor $x=(\Omega GM)^{2/3}$, making its first term formally of 1PN order. This contrasts with Eq. (\ref{emc2}), whose ``first-in-line" term is effectively 0PN, i.e., Newtonian in character. 

If one rewrites Eq. (\ref{emc2}) in an PN-expanded form analogous to Eq. (\ref{enpn2}), thereby making any potential PN energy shifts explicit, we may introduce a modified PN variable for the shell's axial rotation: $\tilde{x}=(a_0\Omega G\mu)^{2/3}$ with $a_0=\sqrt{5.276}$, and express Eq. (\ref{emc2}) in terms of the shell radius $\rho=\rho(t)$ and the speed ratio $\beta=\rho\Omega$. At any arbitrary time $t$ and, in particular, at coalescence $t=t_C$, we obtain for the mass-shell representation:
\begin{eqnarray}\label{epn1}
\Rightarrow\quad &E\simeq1.652\left(\frac{G\mu^2}{\rho}-\frac{\rho}{G}\tilde{x}^3\right)-\sum_{n\in\mathbb{R}}\Delta E_{n\mathrm{PN}},\\\label{epn2}
&E(t_C)\simeq1.652\left(\frac{1}{2}\alpha\mu -2M\tilde{x}_C^3\right)-\sum_{n\in\mathbb{R}}\Delta E_{n\mathrm{PN}}.
\end{eqnarray}
Eq. (\ref{epn2}) closely mirrors Eq. (\ref{emc2}), as well as Eq. (\ref{enpn2}). Since $\tilde{x}_C=(a_0\Omega_CG\mu)^{2/3}=(a_0\alpha\beta_C/2)^{2/3}$, it follows that $2M\tilde{x}^3_C=a_0^2\alpha\mu\beta_C^2/2$, which precisely recovers the quadratic spin term appearing in Eqs. (\ref{energy}) and (\ref{emc2}).

Thus, Eqs. (\ref{epn1}) and (\ref{epn2}) show that Eq. (\ref{energy}) already contains, in compact form, both the 0PN Newtonian term and the 3PN spin-related correction, in agreement with the PN structures identified in Eqs. (\ref{phase}) and (\ref{enpn2}). In the standard PN framework, the 3PN term encodes the axial spins of the CCB masses, whereas here the corresponding contribution is generated by the axial rotation of the mass shell itself -- consistent with the Laplace-Beltrami formalism and the Kerr metric Ansatz underlying this construction. Therefore, the explicit PN-corrected energy shifts $\Delta E_{n\mathrm{PN}}$ in Eq. (\ref{epn2}) now allow one to incorporate effective corrections through powers of $\tilde{x}_C=(a_0\alpha \beta_C/2)^{2/3}$, appropriately scaled by the mass-energy relation $3.304M$. 

\subsubsection{On Eccentricity-Led Corrections}

The motivation of introducing an energy-shift correction originates from the previous discussion in Ref. \cite{MacKay:2024qxj} concerning GW190521; see also Table \ref{t2}. For GW190521, the uncorrected approximation via Eq. (\ref{energy}) gives $6.73 M_\odot$ , whereas the total-minus-remnant difference yields $5.7 M_\odot$. This corresponds to an energy shift of $\Delta E\simeq1 M_\odot$, and it is well established that GW190521 is a highly eccentric event with $e\approx0.7$ \cite{Gayathri:2020coq}. It is therefore \textit{not} an ad hoc assumption that a near-unity eccentricity should lead to a systematic energy correction, as eccentricity already appears as the 2PN correction in the IMR orbital phase shift \cite{Favata:2013rwa} (1.5PN in Ref. \cite{McMillin:2025hof}). 

Adopting the 1.5PN order placement for eccentricity, the corresponding energy shift $\Delta E_\mathrm{1.5PN}$ inferred from Eq. (\ref{epn2}) yields
\begin{equation}
\Delta E_\mathrm{1.5PN}=A_{1.5}\cdot3.304M \left(\frac{a_0}{2}\alpha\beta_C\right)=A_{1.5}\cdot3.795\mu \beta_C,
\end{equation}
where $A_{1.5}$ is a dimensionless PN-specific proportionality coefficient. Using GW190521 as a representative example for this eccentric energy shift, we numerically find the proportionality coefficient:
\begin{equation}
1M_\odot \simeq \Delta E_\mathrm{1.5PN}\Big|_\mathrm{GW190521}~\Rightarrow\quad A_{1.5}\simeq0.0948.
\end{equation}
It is reasonable -- at least as an organizing principle -- to expect that $A_{1.5}$ is gauged by the eccentricity: $A_{1.5}=A_{1.5}(e)$, whereby the current value of $A_{1.5}\simeq0.0948$ is set by $e\approx0.7$ for GW190521.

In standard PN theory, the eccentricity-induced correction to the GW phase shift, entering as e.g. 2PN in Eq. (\ref{phase}), takes the analytical form of a cumbersome linear combination that scales with $2355 e^2/(1462)$ \cite{Favata:2013rwa}. In the energetics picture, the energy flux of GWs generated by eccentric orbits scales with the well-known polynomial $f(e)=(1+73e^2/24+37e^4/96)\cdot(1-e^2)^{-7/2}$ \cite{Blanchet:2013haa}, which modulates the luminosity of GWs generated by eccentric binaries. For the 1.5PN energy shift, it is therefore natural to adopt the phenomenological identification $A_{1.5}=a_1f(e)$, where $a_1$ is a dimensionless fitting parameter relating the calculated $A_{1.5}$ coefficient to the $f(e)$ function under a known $e$., i.e. $e\approx0.7$ that is associated with GW190521. The corresponding fit yields $a_1\simeq0.00348$, or as a fraction $a_1\approx87/25000$. While GW190521 serves a representative, high-eccentricity case, the universality of $a_1$ remains an open question. Testing whether $a_1\simeq0.00348$ is characteristic of all eccentricity-rich events would require examining additional sources with reliably measured eccentricity values. For now, we may claim that
\begin{equation}
\Delta E_\mathrm{1.5PN}\simeq 0.0132\cdot f(e) \mu \beta_C.
\end{equation}

\subsubsection{On Tidal Deformability-Led Corrections}

For tidally induced energy corrections, we can adopt the approach similar to that used for eccentricity, as tidal deformabilities enter the PN expansion on the IMR phase shift at 6PN order \cite{Favata:2013rwa, Flanagan:2007ix, Chatziioannou:2020pqz}. Our primary example of a tidal-deformability-rich GW event is GW170817 \cite{LIGOScientific:2017vwq}, in which the source BNSs each possess a dimensionless tidal deformability parameter $\Lambda_1,~\Lambda_2$, and the binary as a whole is characterized by the combined parameter $\tilde{\Lambda}$. The tidal deformability for each NS takes the form $\Lambda=2k_2C^{-5}/3$, which depends on the $l=2$ Love number $k_2$ and the compactness $C:=Gm/R$ with $R$ being the NS radius \cite{Hinderer:2007mb}.

 It should be noted for completeness that $\Lambda\propto k_2$ is sensitive to one's choice of NS equation of state; for instance, a polytropic form $p=K_i\epsilon^{\Gamma_i}$ may be inserted in the Tolman-Oppenheimer-Volkoff equations and their spherical-harmonic-gauged tidal perturbations. This procedure allows one to extract $k_2$, and thus $\Lambda$, numerically, and various numerical results depend on the chosen NS equation of state. In this section, we take the atheistic approach that does not consider any choice of NS equation of state, as we acknowledge that each NS tidal deformability (much like each NS mass) is strongly spin-prior dependent. E.g., for the larger NS mass with (low-spin prior) center value $m_1=1.46 M_\odot$, the tidal deformability reads as $\Lambda\leq800$ for low-spin priors and $\leq1400$ for high-spin priors \cite{LIGOScientific:2017vwq}. 

However, Table \ref{t1} shows that the GW energy for GW170817 is not well constrained, and the \textit{GWOSC} open source catalog quotes the final mass as $M_\mathrm{f}\leq2.8M_\odot$. If we take this upper bound as a fiducial value, i.e. set $M_\mathrm{f}=2.8M_\odot$, then the radiated energy is directly determined by the total-minus-remnant mass difference. Using our well known binary mass values: $M\equiv m_1+m_2=2.73M_\odot$ via the central values. This immediately introduces a problem: computing $M-M_\mathrm{f}$ with $M_\mathrm{f}=2.8M_\odot$ yields a negative mass difference, and hence an unphysically negative radiated energy for the GW. In such a circumstance, one might resort to the ad hoc correlation between the radiated energy and one-tenth of the chirp mass $E\sim\mathcal{M}/10$, rather than the fractional proportionality $E\propto \mathcal{M}^{5/3}$, to provide a reasonable energy constraint. Using the spin-prior-mutual chirp mass $\mathcal{M}=1.186^{+0.001}_{-0.001}M_\odot$ \cite{GWOSC, LIGOScientific:2018mvr}, we obtain $E\sim 0.1186 M_\odot c^2$.

When $E\sim0.1186 M_\odot c^2$ is compared with $E(t_C)=0.140 M_\odot c^2$ via Eq. (\ref{energy}), we again find that the approximation exceeds the ``observed" value. Thus, we may apply the energy-shift procedure and quantify $\Delta E_\mathrm{6PN}\simeq0.0214 M_\odot c^2$. It is worth noting that this ``tidal energy shift" is significantly smaller than the previous ``eccentric energy shift" obtained for GW190521. This difference is intuitive: eccentric corrections enter at $1.5\sim2$PN order, whereas (combined) tidal deformabilities appear at 6PN. As in standard PN hierarchy, higher PN order terms contribute more subtly to the IMR phase evolution, and thereby to the dynamic angular frequency. Relating $\Delta E_\mathrm{6PN}$ to the corresponding energy shift $\Delta E_\mathrm{6PN}$ yields
\begin{equation}
\Delta E_\mathrm{6PN}=A_6\cdot3.304M \left(\frac{a_0}{2}\alpha\beta_C\right)^4=A_6\cdot5.748\alpha^3\mu \beta_C^4.
\end{equation}
where, just like $A_{1.5}$ for the eccentric energy shift, $A_6$ is dimensionless PN-specific proportionality coefficient. Specifically for GW170817, this coefficient is gauged: 
\begin{equation}
0.0214M_\odot \simeq \Delta E_\mathrm{6PN}\Big|_\mathrm{GW170817}~\Rightarrow\quad A_6\simeq1.328\times10^{9}.
\end{equation}

For the tidal 6PN correction, both the phasing and energetic contributions scale linearly with the dimensionless combined tidal deformability $\tilde{\Lambda}$ \cite{Blanchet:2013haa}. For GW170817, the low-spin prior value is $\tilde{\Lambda}\leq800$ \cite{LIGOScientific:2017vwq} (as low-spin prior mass values were used in Table \ref{t1}; consistency is required). However, the empirically inferred value of $A_6$ for GW170817 is many orders of magnitude larger than $\tilde{\Lambda}$; this makes the faithful linear identification $A_6=a_2\tilde{\Lambda}$ uninformative, as $a_2$ would simply absorb almost the entire numerical scale of $A_6$. To retain physical intuition, we heuristically explore the possibility that the effective tidal dependence is \textit{nonlinear}, adopting e.g. $A_6=a_2\tilde{\Lambda}^m$ with $m\in\mathbb{N}$ to avoid any nonsensical fractional powers. Choosing e.g. $m=3$ provides a simple reconciliation: $A_6=a_2\tilde{\Lambda}^3\quad\Rightarrow \quad a_2\simeq2.594$ for $\tilde{\Lambda}=800$. 

The choice $m=3$ may at first appear ad hoc, but it aligns with the observation that the tidal term contributes at 6PN order. Decomposing the PN order as $6=3\times2$ motivates the identification $\tilde{\Lambda}^3=\tilde{\Lambda}^{6/2}$, which provides a natural whole-integer exponent. Whether $a_2\simeq2.594\approx2.6$ is unique to GW170817 or is characteristic of other tidal-deformability-rich events can only be determined once more BNS systems with well-measured combined deformabilities are observed -- e.g. by next-generation detectors such as the Einstein Telescope (ET). For now, we may claim that
\begin{equation}
\Delta E_\mathrm{6PN}\simeq \frac{299}{20}\tilde{\Lambda}^3\alpha^3\mu \beta_C^4,
\end{equation}
and assuming the specified forms for the 1.5PN and 6PN energy shifts are generalizations, we may write the PN-expanded version of Eq. (\ref{energy}), extending Eq. (\ref{epn2}) to include the fitted eccentric and tidal contributions explicitly:
\begin{eqnarray}\nonumber
E(t_C)\simeq&~1.652\left[\frac{1}{2}\alpha\mu -2M\tilde{x}_C^3\left(1+\frac{87}{25000}f(e)\tilde{x}_C^{-3/2}+\frac{13}{5}\tilde{\Lambda}^3\tilde{x}_C^3\right)\right]\\
&-\sum_{n\in\mathbb{R}}\Delta E_{n\mathrm{PN}}.
\end{eqnarray}

\section{Concluding Statements}\label{concl}

In this work, we reinterpret the EOB framework \cite{Buonanno:1998gg, Buonanno:2005xu} through the rotating and contracting CCB mass-shell model introduced in Ref. \cite{MacKay:2024qxj}. This model provides an opportunity to approach the EFEs in a novel manner: via a variational method applied to the Laplace-Beltrami formalism with a well-defined metric Ansatz, e.g. Kerr for a spinning, compact configuration. Within this framework, we obtain an approximate expression for the radiated quadrupolar GW energy as the corresponding energy (Eq. [\ref{energy}]) to an effective $T_{00}$ calculation. This expression yields predictions that are, for the most part, in close agreement with observations as demonstrated for several representative events and summarized in Tables \ref{t1}--\ref{t4} and reinforced in Section \ref{goods}. Residual discrepancies -- even in otherwise well-matched cases -- can be interpreted as PN-corrected energy shifts, as discussed throughout Section \ref{coll}. This motivates the PN-expanded form of the coalescence energy (Eq. [\ref{epn2}]), which makes such contributions explicit and allows eccentricity, tidal deformability, and other higher-order effects to be incorporated in a unified energetics picture.

Future extensions of this work include further comparisons with the remaining, unaccounted cataloged GW events, refining the PN-correction terms in this effective energetics picture -- exploring possible ``fine-tuning" of the correction structure introduced here --, and analyzing the mass-shell waveforms generally via Eq. (\ref{pieces}) or specifically via Eqs. (\ref{hp2}) and (\ref{hc2}) when applied to certain (relevant) inspiral-merger stages.  

It is also of interest to analyze the broader application of the Laplace-Beltrami formalism beyond the CCB mass-shell model. Examining alternative metric Ans\"atze may further clarify both the utility and the limitations of this variational approach to the EFEs.

%\pagebreak

\section*{Acknowledgments}

I thank the referees on their insightful comments, which have further enhanced this work.

\section*{Statement Declarations}

\subsection*{Conflict of Interest}
The author declares no conflicts of interest.

\subsection*{Data Access Statement}
As a theoretical study, this work generates no original data. Data from cited LIGO observations are publicly available.

\subsection*{Ethics Statement}
No ethical issues arise, as no test subjects are involved. This paper adheres to academic integrity.

\subsection*{Funding Statement}
This work received no funding.

%\appendix


% Please avoid comments such as "For a review'', "For some examples",
% "and references therein" or move them in the text. In general,
% please leave only references in the bibliography and move all
% accessory text in footnotes.

% Also, please have only one work for each \bibitem.

\end{document}